\documentstyle[12pt,aasms4]{article}

\newread\epsffilein    
\newif\ifepsffileok    
\newif\ifepsfbbfound   
\newif\ifepsfverbose   
\newdimen\epsfxsize    
\newdimen\epsfysize    
\newdimen\epsftsize    
\newdimen\epsfrsize    
\newdimen\epsftmp      
\newdimen\pspoints     
\def\epsfbox#1{\global\def\epsfllx{72}\global\def\epsflly{72}%
   \global\def\epsfurx{540}\global\def\epsfury{720}%
   \def\lbracket{[}\def\testit{#1}\ifx\testit\lbracket
   \let\next=\epsfgetlitbb\else\let\next=\epsfnormal\fi\next{#1}}%
\def\epsfgetlitbb#1#2 #3 #4 #5]#6{\epsfgrab #2 #3 #4 #5 .\\%
   \epsfsetgraph{#6}}%
\def\epsfnormal#1{\epsfgetbb{#1}\epsfsetgraph{#1}}%
\def\epsfgetbb#1{%
\openin\epsffilein=#1
\ifeof\epsffilein\errmessage{I couldn't open #1, will ignore it}\else
   {\epsffileoktrue \chardef\other=12
    \def\do##1{\catcode`##1=\other}\dospecials \catcode`\ =10
    \loop
       \read\epsffilein to \epsffileline
       \ifeof\epsffilein\epsffileokfalse\else
          \expandafter\epsfaux\epsffileline:. \\%
       \fi
   \ifepsffileok\repeat
   \ifepsfbbfound\else
    \ifepsfverbose\message{No bounding box comment in #1; using defaults}\fi\fi
   }\closein\epsffilein\fi}%
\def\epsfsetgraph#1{%
   \epsfrsize=\epsfury\pspoints
   \advance\epsfrsize by-\epsflly\pspoints
   \epsftsize=\epsfurx\pspoints
   \advance\epsftsize by-\epsfllx\pspoints
   \epsfxsize\epsfsize\epsftsize\epsfrsize
   \ifnum\epsfxsize=0 \ifnum\epsfysize=0
      \epsfxsize=\epsftsize \epsfysize=\epsfrsize
     \else\epsftmp=\epsftsize \divide\epsftmp\epsfrsize
       \epsfxsize=\epsfysize \multiply\epsfxsize\epsftmp
       \multiply\epsftmp\epsfrsize \advance\epsftsize-\epsftmp
       \epsftmp=\epsfysize
       \loop \advance\epsftsize\epsftsize \divide\epsftmp 2
       \ifnum\epsftmp>0
          \ifnum\epsftsize<\epsfrsize\else
             \advance\epsftsize-\epsfrsize \advance\epsfxsize\epsftmp \fi
       \repeat
     \fi
   \else\epsftmp=\epsfrsize \divide\epsftmp\epsftsize
     \epsfysize=\epsfxsize \multiply\epsfysize\epsftmp   
     \multiply\epsftmp\epsftsize \advance\epsfrsize-\epsftmp
     \epsftmp=\epsfxsize
     \loop \advance\epsfrsize\epsfrsize \divide\epsftmp 2
     \ifnum\epsftmp>0
        \ifnum\epsfrsize<\epsftsize\else
           \advance\epsfrsize-\epsftsize \advance\epsfysize\epsftmp \fi
     \repeat     
   \fi
   \ifepsfverbose\message{#1: width=\the\epsfxsize, height=\the\epsfysize}\fi
   \epsftmp=10\epsfxsize \divide\epsftmp\pspoints
   \vbox to\epsfysize{\vfil\hbox to\epsfxsize{%
      \includegraphics{#1}%
      \hfil}}%
\epsfxsize=0pt\epsfysize=0pt}%

{\catcode`\%=12 \global\let\epsfpercent=
\long\def\epsfaux#1#2:#3\\{\ifx#1\epsfpercent
   \def\testit{#2}\ifx\testit\epsfbblit
      \epsfgrab #3 . . . \\%
      \epsffileokfalse
      \global\epsfbbfoundtrue
   \fi\else\ifx#1\par\else\epsffileokfalse\fi\fi}%
\def\epsfgrab #1 #2 #3 #4 #5\\{%
   \global\def\epsfllx{#1}\ifx\epsfllx\empty
      \epsfgrab #2 #3 #4 #5 .\\\else
   \global\def\epsflly{#2}%
   \global\def\epsfurx{#3}\global\def\epsfury{#4}\fi}%
\def\epsfsize#1#2{\epsfxsize}
\def\lapprox{\hbox{\lower .8ex\hbox{$\,\buildrel < \over\sim\,$}}}
\def\gapprox{\hbox{\lower .8ex\hbox{$\,\buildrel > \over\sim\,$}}}
 
\begin{document}

\title{Bolometric light curves of supernovae and post--explosion
 magnetic fields}

\author{P. Ruiz--Lapuente \altaffilmark{1,2}, and  H. C. Spruit
\altaffilmark {2}}

\altaffiltext{1} {Department of Astronomy, University of Barcelona
Mart\'\i\ i  Franqu\'es, 1 -- E--08028 Barcelona, Spain}

\altaffiltext{2} {Max--Planck--Institut f\"ur Astrophysik
Karl--Schwarschild--Str. 1 -- D--85740 Garching, Germany}

\slugcomment{{\it Running title: } SNe light curves and magnetic fields}

\abstract{ The various effects leading to diversity in the bolometric light
curves of supernovae are examined: nucleosynthesis, kinematic differences,
ejected mass, degree of mixing, and configuration and intensity of the
magnetic field are discussed. In Type Ia supernovae, a departure in the
bolometric light curve from the full--trapping decline of $^{56}$Co can occur
within the two and a half years after the explosion, depending on the
evolutionary path followed by the WD during the accretion phase. If
convection has developed in the WD core during the pre--supernova evolution,
starting several thousand years before the explosion, a tangled magnetic
field close to the equipartition value should have grown in the WD. Such an
intense magnetic field would confine positrons  where they originate from the
$^{56}$Co decays, and preclude a strong departure from the full--trapping
decline, as the supernova expands. This situation is  expected to occur in
C+O Chandrasekhar WDs as opposed to edge--lit detonated  sub--Chandrasekhar
WDs. If the pre--explosion magnetic field of the WD is less intense than
10$^{5-8}$G, a lack of confinement of the positrons emitted in the $^{56}$Co
decay and a departure from  full--trapping of their energy  would occur. The
time at which the departure takes place can provide estimates of the original
magnetic field of the WD, its configuration, and also of  the mass of the
supernova ejecta. In SN 1991bg, the bolometric light curve suggests absence
of a significant tangled magnetic field: its intensity is estimated to be
lower than  $10^{3}$ G. Chandrasekhar--mass models do not reproduce the
bolometric light curve of this supernova.    For SN 1972E, on the contrary,
there is evidence for a tangled  configuration of the magnetic field and its
light curve is well reproduced by a Chandrasekhar WD explosion. A comparison
is made for the diagram  of absolute magnitude and rate of decline in Type Ia
supernovae coming from different explosion mechanisms.   The effects of
mixing and ejected mass in the bolometric light curve of Type Ibc supernovae
are also discussed.}


\section{ Introduction}

 Light curves of supernovae vary significantly. Even within
the same supernova type, a spread in the shape of the light curves
 is found. So far, the differences have not been quantified 
in terms of a
spread in the integrated bolometric light curves, but rather in terms 
of the light curves
in the various broad--band filters (Hamuy et al. 1996a,b; Riess, Press
 \& Kirshner 1996).
   
From the theoretical point of view, several factors can induce changes in the
evolution in luminosity within a sample of supernovae of the same type:
different distributions of radioactive material in velocity space resulting
from differences in the burning propagation along the star, a spread in total
masses of the ejecta, or a diversity in the configuration and intensity
of the magnetic field, B, of the stars prior to explosion.

Very few studies have been done on the evolution of the magnetic field of a
star which explodes, and how this affects its overall luminosity.
In particular, little
attention has been devoted to the fate of the original magnetic field
configuration, which should experience a drastic change due to the enormous
expansion undergone by the supernova ejecta. That can bear observable 
consequences in the evolution of the luminosity of the supernova.

The progenitor stars of thermonuclear supernovae are appreciably magnetized
objects. Magnetic fields of WDs have been measured and 
range from 10$^{5}$ to as much as 5 $\times$ 10$^{8}$ G (Liebert 1995). 
Prior to
the explosion, the turbulent motions inside the WD can  alter the original
intensity and configuration of such field by fast dynamo action. After the
explosion, the huge expansion undergone by the ejecta reduces 
the magnetic field inside
the supernova. The evolution with time of the supernova field 
becomes relevant to the trapping of the
energy of the positrons originated in the 
radioactive $\beta^{+}$--decays of $^{56}$Co.
Several hundreds days after the explosion,  if the magnetic
field lines do not contribute to confine the path of the energetic positrons, 
with kinetic energies in the MeV range, 
a fraction of this energy escapes
the innermost ejecta and the evolution in luminosity of  
the supernova is affected.

Colgate, Petschek, \& Kriese (1980), and more recently Colgate (1991,1997)
undertook a determination of the ejected mass in supernovae through the
escape of $\beta^{+}$ energy in the tail of the light curves. Significant
departures from the $^{56}$Co--decay full--trapping curve are argued
 in those
 works.  Axelrod (1980), on the other hand, suggested in his SN modeling
that a chaotic weak magnetic field of B $\approx$ 10$^{-6}$ G 
after the explosion would confine the
positrons up to late phases. Under the later assumption, 
the late decline in
luminosity approaches the $^{56}$Co--decay full--trapping line, although
positron energy is not fully deposited (Chan \& Lingenfelter 1993).
Whereas in earlier works a particular configuration of the magnetic 
field has been assumed, in the
present work, we look at the processes undergone by the WD prior to
explosion and as it expands, and predict how the magnetic field contents
might evolve. The suggested evolution should then be compared with the
observations through the predictions of the supernova luminosity. 

In thermonuclear supernovae ( i.e. Type Ia supernovae, or SNe Ia), depending
on whether the diversity among the bolometric tails is moderate (of the order
of 10---15$\%$) or larger ($\ge$ 30--50 $\%$) in the fraction of energy
deposited, one would favor different mechanisms to explain
 this diversity. As will be shown in this work, moderate
diversity suggests different distributions of radioactive material
in velocity space, and a larger diversity implies differences in the magnetic
field in the ejecta, and possibly also
differences in the ejected mass. The information on the post--explosion
magnetic field derived from the SN late luminosity can be linked to the
nature of the pre--explosion magnetic field.

In the case of Type Ibc supernovae, the effect of mixing in the deposition of
energy from both $\gamma$--rays and positrons is examined as an important
factor in determining the final shape of their light curves.

In Sections  2 and 3, $\gamma$--ray  and $\beta^{+}$--energy deposition
calculations of supernovae are presented. In section 4, it is shown how in
 SNe Ia both mass and nucleosynthetic distribution as well as pre and
post--explosion magnetic field, can be investigated through the study of the
bolometric light curves. The present understanding of the evolutionary stages
previous to explosion, and the duration of processes such as turbulent
convection in accreting WDs, are examined in order to establish the changes
 of the WD magnetic field. In Section 5, the diversity of 
bolometric
declines is presented in terms of the physical processes from which 
it can originate.
The influence of different nucleosynthesis and kinematics of SNe Ia 
in the final luminosity is discussed. Mixing in
supernovae of core--collapse with small ejected mass such as Type Ibc, is
discussed for its effects on the bolometric luminosity within 
the  whole core--collapse SN class as well as its incidence on
 the derivation of the ejected mass  for those low-mass ending
stars. Finally, in Section 6, we infer from the
study of positron escape in supernovae some consequences for 
the origin of the 511 keV positron annihilation line in our Galaxy.

\section{ Radioactivity from $^{56}$Co: $\gamma$--rays and positrons}

In supernovae, the radioactive decay $$^{56}Co \rightarrow  ^{56}Fe$$
provides the source of luminosity along the tail of the light curve. Such a
decay has a half life of 77 days and 81$\%$ of the time gives rise to a
$\gamma$--ray photon and 19$\%$ to a e$^{+}$. $\gamma$--ray photons are
emitted with a spectrum of energies reaching up to 1.4 MeV, and carry about
96.5 $\%$ of the energy of the $^{56}$Co decay. The emitted positrons have an
energy spectrum extending up to the endpoint kinetic energy E$_{max}$=1.459
MeV, and they account for 3.5$\%$ of the energy of the $^{56}$Co decay.  The
fate of this 3.5 $\%$ of energy is crucial at late times.

Compton scattering of the emitted $\gamma$--rays is the main process
degrading the energy of the photons as it is transferred to the electrons  of
the gas which become nonthermal. The comparative simplicity of the process
degrading the energy of the $\gamma$-rays in expanding ejecta allows us to
calculate accurately the energy deposited and the escape of energy as well.
 
Transport calculations of the $\gamma$-rays provide the fraction of
radioactive energy deposited in the supernova ejecta as a function of inner
mass fraction and time. The deposition function $D_{\gamma}(t)$ is a 
decreasing function of time as the supernova expands. The final injection
of energy in the supernova ejecta takes place at a rate:

$$\xi(t) = (6.76\times10^{9}\ D_{\gamma}(t) + 2.72\times10^{8}\ D_{\beta}
(t)) \left(e^{-t/\tau_{Co}} - e^{-t/\tau_{Ni}}\right)$$
$$\hskip 6.2 true cm +\ 3.91\times10^{9}\ D_{\gamma}(t) e^{-t/\tau_{Ni}}\
\rm {erg\ g^{-1}\ s^{-1}}\eqno(1)$$
          
\noindent
where $\tau_{Ni}$=8.8 days  and $\tau_{Co}$=111.26 days are the e--folding
times for radioactive decay of Ni and Co respectively, and D$_{\beta}$ is the
deposition function of  e$^{+}$ energy, whose importance becomes crucial as
the ejecta become transparent to $\gamma$--rays. The term related to the
decay $^{56}Ni \rightarrow  ^{56}Co$ is relevant for the early rise to
maximum luminosity.

Once $\gamma$--ray photons suffer Compton scattering either they do not lose
a significant amount of energy (forward scattering), or they lose
significantly their energy, becoming unable to produce further energetic 
electrons. This has suggested  the adequacy of treating the Compton
scattering process as an absorption process, for applications related to the
energy deposition of $\gamma$--rays (Sutherland \& Wheeler 1984). A similar
approach to that developed by those authors is used to calculate
$\gamma$--ray transport in this work. Two methods of calculation of the
deposition of energy were previously compared: the ``absorption'' approach
generalized for an arbitrary $^{56}$Ni distribution was tested against
detailed Monte Carlo calculations. As found by previous authors (Swartz,
Sutherland, \& Harkness 1995) both results gave a very similar deposition
function. Background models such as the W7 model by  Nomoto, Thielemann, \&
Yokoi (1984)  were used for these tests.

As one follows the evolution of the bolometric light curve of SNe along the
$^{56}$Co tail, different phases can be outlined. For SNe Ia, the
post--maximum decline of the light curve is primarily determined by the
temporal evolution of $ D_{\gamma}(t)$. The luminosity at that phase and its
rate of decline are related to the degree of escape and deposition from those
energetic photons. That degree depends on the distribution of $^{56}$Ni in
the velocity--mass  space, and on the total optical depth of the ejecta. This
suggests defining a $\Delta m_{\gamma}^{100}$ as the number of bolometric
magnitudes of decline per day  during the phase when $\gamma$--rays are the
main contributors.

Later on, D$_{\gamma}$ falls below the contribution of energy  by positrons.
At that time a new inflection in the bolometric light curve shape occurs
linked to the slower evolution in time of $D_{\beta}(t)$. The steepness of
the decline is then related to the distribution of the radioactive $^{56}$Co,
the velocity structure, and to the intensity and configuration of the
magnetic field.  As we will see, different behaviors are expected and they
can give clues to the mechanism of explosion. What happens in the phase when
positrons are the dominant luminosity source depends very much on B.

When positrons in supernovae start to play a major role in the energy input
(as soon as $\tau_{\gamma}$ becomes very small), the fraction of  escape and
deposition of the kinetic energy of those particles establishes the
luminosity. The most energetic positrons and those emitted in the outer
layers may succeed  escaping the ejecta without becoming thermalized. 
 A numerical evaluation is required once the supernova
physical properties are known.
The energy spectrum of the positrons covers a broad range of energies with a
distribution of the form:

$$S(\epsilon)\propto F(Z, \epsilon) (\epsilon_{0} - \epsilon)^{2} \epsilon
\sqrt{\epsilon^{2} - 1}\eqno(2)$$

\noindent
where $\epsilon$ is the total positron energy in units of
 $m_{e}c^{2}$; $ \epsilon_{0}
= E_{max}/m_{e}c^{2} + 1$, $E_{max}$ being the maximum kinetic energy; and
$F(Z,\epsilon)$ is a correction for the Coulomb interaction with the final
nucleus of electric charge Z (Segr\'e 1977):

 $$F(Z, \epsilon) = {2\pi\xi\over 1 - {\rm exp}(-2\xi)}\eqno(3)$$    

\noindent
with
 
$$\xi = -{Ze^{2}\over \hbar v} = -{Z\alpha\over
\sqrt{1-\epsilon^{-2}}}\eqno(4)$$

\noindent
where $Z = 26$, $v$ is the speed of the positron, and $\alpha$ is the
fine--structure constant (Segr\'e 1977).

Positrons with $\beta$ as large as 0.94 ($\beta = v/c$) are produced in the
decay.  Given the initial range of kinetic energies -- in the keV and MeV
range--, the positrons slow down in the supernova ejecta mainly by ionization
and excitation losses. At higher energies, bremsstrahlung would be the
dominant energy loss mechanism, and at lower energies, Coulomb scattering
would be dominant (see Segr\'e 1977, for instance).  

As the positrons slow down due to their loss of energy in the ejecta, they
travel a fraction of the envelope which can be estimated as the {\it stopping
distance, $d_{e}$} due to ionizations and excitations in the SN Ia envelope.

The full relativistic expression for positron energy loss, per unit length,
X, due to ionization of atoms is (Heitler 1954; Blumenthal \& Gould 1970;
Gould 1972):

$${dE\over dX} = - \Gamma (E) = - {4\pi r_{0}^{2}m_{e}c^{2}Z \rho \over
\beta^{2}Am_{n}} {ln\left({\sqrt{\gamma - 1} \gamma \beta \over
I/m_{e}c^{2}}\right) + {1\over 2} ln \ 2 + \Sigma_{2}(E)}\eqno(5)$$

\noindent
where E is the kinetic energy, $r_{0}$ is the classical electron radius,
$m_{n}$ is the atomic mass unit, $Z$ and $A$ are, respectively, the effective
nuclear charge and atomic mass of the ejecta material, and $\Sigma_{2} (E)$
gives the relativistic factors as a function of the Lorentz factor, $\gamma$,
and of $\beta$  (Berger \& Seltzer 1954).

\noindent
$I$ is the effective ionization potential for the ambient atoms in the
ejecta. A semiempirical formula for the ionization potential gives (Roy \&
Reed 1968; Segr\'e 1977):

$$I = 9.1Z\left(1 + {1.9\over Z^{2/3}}\right) eV\eqno(6)$$

Due to the weak dependence of  $dE/dX$ on $I$, the formula above for I is
accurate enough for the practical calculation of the energy loss.

 The {\it stopping distance of the positron}
as result of impact ionization and
excitation in a SN Ia envelope is found to be approximately:

$$d_{e} \equiv {E\over {-dE/dX}} \approx  {3.36\over \rho} \left({E\over
m_{e}c^{2}}\right){A\over Z}(ln{E\over I} )^{-1}\ cm \eqno(7)$$

In Table 1 some typical values are given for d$_{e}$,  the 
{\it stopping distance
of the positrons} of different energies both for Chandrasekhar WDs
and WDs of the smallest possible exploding mass.
Synchrotron losses by the e$^{+}$ in the presence of the magnetic field,
bremsstrahlung losses, and losses due to Compton scattering off photons
contribute to the slowing down of the positrons to a much lesser extent.

Each magnetic field configuration specifies in a given way  the positron
transport in the supernova ejecta. We have specified three  likely
configurations of the field lines, and adopted an efficient way to calculate
the deposition function. Three situations which the positrons might encounter
 in exploded ejecta are: a {\it chaotic magnetic field background} (a likely
result of the turbulent motions  prior to explosion), a  {\it radial field} 
(resulting from expansion of the original dipole  field in fast moving
ejecta), or {\it the absence of a significant magnetic field}, 
in which case they
are just subjected to their interactions with ions and electrons along free
trajectories.

The deposition calculation consists in determining how efficiently the
relativistic positrons transfer their energy to ions and  electrons
increasing the kinetic energy of the latter: positrons thermalize if 
they release most of their kinetic energy. That energy should reappear 
as optical--infrared luminosity through the excitation of a whole range of
transitions or through ionization and recombination processes.  In the
present work, the confinement of positrons  in a chaotic  magnetic field is 
first investigated.  Positrons of different energies are followed through their
interactions over time, testing whether they become thermal or whether they
remain nonthermal within the ever more diluted ejecta. 
 The positron
mean free path is very small as compared with the characteristic radius of
the supernova ejecta when the density of the ejecta is still
high enough to produce large losses of the positron energy by ionization and
excitation, or  when  the presence of a   turbulent magnetic field inside the
ejecta confines the trajectories of the positrons along the winding field
 lines
and induces a larger number of interactions. The mean free path of the
positron becomes large when either the density of the ejecta is too low to slow
down the positrons or the energy density of the magnetic field is 
extremely low and therefore the Larmor gyroradius of the particle is a
sizeable fraction of the radius of the ejecta. In the latter case the escape 
is enhanced. The distance travelled by the positron increases when a strong
radial  magnetic field confines the positrons to move out in their helical
 motions along the radial field lines.

In the presence of a background chaotic magnetic field, positrons of energy
$E_{i}$ born at a given radius r$_{i}$ (of mass coordinate $m_{i}$ and
velocity $v_{i}$) cannot slow down to thermal energies if they are
emitted after a critical time $t_{i} > t_{c} (m_{i}, E_{i})$.
 The turbulent magnetic field confines the positrons at their site
of origin, but as the ejecta expand and decrease in density, the
possibilities for thermalization decrease. Thus, a fraction of the
 energetic positrons will not
succesfully thermalize even under  confinement
and survive in the ejecta as a ``fast'', nonthermal
population.
 The critical time for
thermalization  depends on the gradient of velocity along the ejecta, on the
energy of the positrons, and on their rate of energy loss. A useful 
expression to evaluate such critical time is given by Chan \& Lingenfelter
(1993):

$$t_{c}(m_{i}, \gamma_{i}) = \left[{8\pi m_{e}cv_{sn}^{2}(m_{i})\over M}
\left({dv_{sn}\over dm}\right)_{m_{i}} \times \int_{1}^{\gamma_{i}}
{\gamma\over \Gamma(\gamma m_{e}c^{2})\sqrt{\gamma^{2} - 1}}\
d\gamma\right]^{-1/2}\eqno(8)$$

\noindent
where M is the mass of the ejecta, $v_{sn}(m_{i})$ is the velocity of the
supernova ejecta at $m_{i}$, $(dv_{sn}/dm)_{m_{i}}$ is the velocity gradient
at the location of $m_{i}$, and $\Gamma(E)$ is the energy loss due to the
different processes. Chan \& Lingenfelter (1993) included among those 
processes impact ionization and excitation, whereas we found that 
one should include as well Coulomb scattering (Bhabha scattering
involving e$^{+}$ e$^{-}$, in this case),
since this is also a major process in the deposition of energy of 
 positrons. Our algorithm differs from that from those authors in the
inclusion of this additional process as degrading the positron kinetic 
energy, and also in the focus of the calculations: the main quantity
for light curve calculations is 
the energy deposited in the supernova, instead of the
energy escaping as energetic positrons.

In the second configuration considered here, the confinement of
 positrons in a
chaotic magnetic field is substituted for a different frame: the particles
travel along the lines of a radial magnetic field.  Again, $\Gamma (E)$, the
energy loss function, and the mass of the ejecta will determine the 
fraction of kinetic energy that they deposit. The equation for the
 trajectory has to be
solved simultaneously with the energy loss equation.

$$r = v_{sn} (m_{i}) \  t_{i} + \int_{t_{i}}^{t} c \beta (t') cos [\theta
(m, t')] dt'\eqno(9)$$
  
\noindent
given an initial mass coordinate $m_{i}$ and pitch angle $\theta_{i}$. The
changes in pitch angles due to the gradient in B(r,t) outwards, favor a
forward beaming of the positrons in the radial direction, even if they were
emitted with $\theta_{i}$ close to $\pi/2$ (Colgate, Petschek, \& Kriese 1980;
Chan \& Lingenfelter 1993).

A last option is the
absence of any significant magnetic field able to affect the trajectory of
the particle. In that case, positrons are not confined to follow 
any trajectories and a treatment similar to $\gamma$--ray transport 
can  be used, adopting the appropiate absorption coefficient for the 
positron processes.

The default values given in the Tables for bolometric magnitude declines
correspond to the chaotic field case, but decline rates in the absence of
magnetic field and in the case of a radial field will also be mentioned
when comparing models with observations.

\section{ Time evolution of the bolometric magnitude}

\subsection{ C+O WD ignition and structure of the radioactive source }

The degree to which differences observed in 
the evolution in luminosity of supernovae are linked to the distribution 
of radioactive sources and the kinematic structure of the exploded star has
hardly been quantified. 
The central ignition of a C+O WD with a mass close to the Chandrasekhar mass
produces a $^{56}$Ni distribution buried from the center up to variable
mass fractions, depending on the characteristics of the burning front.  
 In the alternative  edge--lit 
detonations of C+O WDs, 
burning starts at the outermost layers of the star and proceeds towards
the center. In those  explosions,  the radioactive material is found at two
different locations: very near to the surface, where the ignition started, 
and around the center where, after propagation of the burning front,
the densities  of the interior favor burning to NSE (Nuclear Statistic
Equilibrium) products. 

  Within both frames for the explosion of a SNIa,
variations in the total mass of $^{56}$Ni and of its location in velocity
space are found, related to the extent to which the burning front incinerates
the material.  Classical Chandrasekhar central ignition models are able to
incinerate 0.6 M$\sb\odot$ of the star to $^{56}$Ni.  The nucleosynthesis
 and density structure of the
class is well represented by model W7
 (Nomoto, Thielemann, \& Yokoi 1984), where 0.63 M$\sb\odot$ of
$^{56}$Ni are buried below the surface of 9000 km s$^{-1}$. This model is
known to provide a good spectrum for ``normal SNe Ia''.  In the case of
Chandrasekhar explosions, very $^{56}$Ni--poor explosions can also be found
when the WD undergoes a pulsation that changes  the mode of propagation of
the burning front (Khokhlov 1991).  The pulsating delayed detonations can
produce a low amount of $^{56}$Ni in the center, which is buried  at low
velocities and very low mass fractions. An example of such model  is the here
depicted  model WPD1 (Woosley 1997), suggested to account for subluminous SNe
Ia.

Among sub--Chandrasekhar models, a range of possible explosions corresponds
to the ignition of WDs of different masses (Woosley \& Weaver 1994; Livne \&
Arnett 1995). Results from 1--D and 2--D hydrodynamic calculations give
similar final structures for the ejecta, and a whole range of possible
structures corresponding to the ignition of WDs of different masses. An
exploded WD of mass $\simeq$ 0.97 M$\sb\odot$ synthesizes the same amount of
$^{56}$Ni as W7, but it contains this radioactive element also  in the
outermost layers (model 6 by Livne \& Arnett 1995, for instance).  The
detonation of a 0.7 M$\sb\odot$ C+O WD synthesizing about 0.15 M$\sb\odot$ of
$^{56}$Ni corresponds to the  lowest end of possible WD masses able to
explode by edge-lit detonations (model 2 by Livne \& Arnett 1995
represents such a structure). It is a candidate to explain very subluminous SNe
Ia. On the highest end in luminosity, the detonation of a 1 M$\sb\odot$ C+O
WD provides the largest amount of $^{56}$Ni ($\simeq$ 0.97 M$\sb\odot$). As
representative of the top end,  we investigate a model by Nomoto (1994).
Table 2 summarizes the characteristics of the models investigated here as
possible structures of exploded WDs.

\subsection{ D$_{\gamma}$(r) and $\Delta {m_{\gamma}^{100}}$ in Type
Ia supernovae }

The bolometric light curve in the phase where the $\gamma$--rays fuel the
luminosity is well described by the decrease between one hundred and 
two hundred days after
the explosion. We can define $M_{bol}^{100}$ as the bolometric magnitude at
100 days after explosion and  $\Delta {m_{\gamma}^{100}}$ {\it as the number of
bolometric magnitudes declined per day after 100 days}.  During the period
where $\Delta {m_{\gamma}^{100}}$ measures the decline rate
 due to the increasing transparency of
the envelope to $\gamma$--rays, the SNIa models of lower mass do not
experience a larger change of magnitude  than the more massive ejecta.
This is due to the fact that the $\gamma$--rays of the outermost $^{56}$Ni
have always escaped easily, and, on energy deposition effects they where not
important. The inner structure of the deposition function shows a peak  at
the innermost radii as in the Chandrasekhar models, and the  density
structure is somewhat flatter in less massive WDs.  The luminosity of
Chandrasekhar models is, however, higher  than that of low--mass models
because the total $^{56}$Ni mass being equal, the effectively buried
 $^{56}$Ni relevant for $\gamma$--ray trapping is larger
 than in the edge-lit cases, and
the optical depth is larger (more mass).  Table 3 gives the values for
 $\Delta {m_{\gamma}^{100}}$, and absolute magnitudes for various
models.

Figure 1 shows by the example of subluminous SNe Ia resulting from the 
sub--Chandrasekhar explosion of a 0.7 M$\sb\odot$ (model 2 by Livne \& Arnett
1995) or from a pulsating delayed detonation (model WPD1 by Woosley 1997), the
level of deposition of $\gamma$--rays of the supernova at late 
phases. The model of pulsating delayed detonation achieves a higher
luminosity than the sub--Chandrasekhar edge--lit detonation of a 0.7
M$\sb\odot$ WD. The Figure also shows  how most of the emission comes from
the inner 20$\%$  fraction of mass.

If $^{56}$Co would only give $\gamma$--rays, we would never see the 50--80
$\%$ of mass fraction in SNe Ia at late phases. Since the
$\gamma$--ray--sphere (if we define it as the sphere which concentrates more
than  80$\%$ of the deposition in energy) has shrunk down to the 20$\%$
inner mass fraction or even deeper, we would only see emission at very
low velocities. Due to the role of positrons and to their flatter deposition
function, this does not actually happen. Positrons stop the drop in
luminosity of the supernova ejecta.

\subsection{Decline in the positron tail}

Leaving aside the discussion on the magnetic field intensity (it will be
addressed in the next section), it is clear that some inner structures of
exploded WDs are more favorable to trapping of positrons than others. The
effect can start to become evident as early as 200 days after the explosion.
For larger velocity gradients, and $^{56}$Ni placed outside the inner
regions, the escape of positron energy is enhanced. The escape strongly
depends on the velocity gradient along the ejecta and on the distribution of
the radioactive source. Better trapping structures  are centrally--ignited
Chandrasekhar C+O WDs, as compared with  edge--lit C+O WDs.  If we determine
the {\it number of magnitudes declined per day in the bolometric light curve
between 200 and 400 days},  $\Delta m_{1 \beta}$, {\it and between  400
days and 1000 days}, $\Delta m_{2 \beta}$, a good clue as to the right
model can be achieved. The expected values for some representative models are
given in Table 3.  Table 3  compares the rate of decline between 200 and 400
days when positrons are the main energy source with that later on, between
400 and 1000 days, when they might fail to fully deposit their energy,
depending on the post--explosion  structure of the supernova. This Table
stresses the intrinsic differences due to the  kinematics and distribution of
radioactive material in different models. The calculations have been done
assuming confinement by B within the ejecta. In Section 5 we relax this
requirement in view of the existing observations.

The difference in deposition of positron energy at 300 days and 350 days for
the abovementioned SNe Ia models, if the magnetic field succeeds in providing
enough trapping in the ejecta, is displayed in Table 4. As it can be seen,
escape in sub--Chandrasekhar WDs can reach up to 20$\%$ after 200 days. 
In Figure 2
the deposition function of the $\beta^{+}$ energy shows different behaviors
for the different models from 200 to 1000 days. In
a Chandrasekhar model like W7 even with a turbulent configuration of the
magnetic field, at 740 days after the explosion the positrons do not deposit
their energy totally. However, the departure is only moderate ($10 \%$).
 If the configuration of the magnetic field were
radial, or the magnetic field intensity were very low,  significant
 departures from
full trapping could be achieved even earlier.

\section { Magnetic field of the WD: pre and post-explosion}

Either full confinement of positrons in a chaotic magnetic field (Axelrod
1980), or a enhanced escape through a radially combed out configuration
(Colgate, Petschek, \& Kriese 1980; Colgate 1991) have been considered as
limiting cases for the magnetic field in the SN ejecta. From 
current knowledge of the pre--explosion structure of the WD, and following
the effects of expansion, we can reexamine these issues.

Some WDs are known to host magnetic fields of intensity ranging between
10$^{5}$ and 10$^{9}$ G (Liebert 1995). Such high magnetic fields are not
common, however. Most WDs probably have fields below the detection limit of
10$^{4}$--10$^{5}$ G. A number of studies suggest  that the configuration of
the magnetic field is generally  more complex than a dipole (non--centered
dipolar geometry or quadripolar), and that the strength of the magnetic field
might be correlated with the mass of the WD in the sense of more massive
WDs hosting larger magnetic fields. This last point, however, has not been
established on firm statistical basis.

Prior to the huge expansion induced by the explosion, however, the 
mass--accreting WD progenitor of the SNe Ia goes through a stage which 
might increase the intensity of its initial magnetic field. 
Thermonuclear runaway, which initiates the explosion, is preceded by a 
stage of quasistatic C burning that creates a central convective core
(Woosley et al. 1990; Niemeyer \& Hillebrandt 1995). As already found
by Arnett (1969), when C burning accelerates after C ignition, electron
conduction alone becomes unable to transport the ever larger energy flux
and the temperature gradient becomes superadiabatic. Therefore, turbulent
motions develop, encompassing a sizeable fraction of the WD interior.
The interaction of the turbulence with the magnetic field should thus
be examined, in order to see whether an initially small $B$ might be 
significantly amplified during the steps immediately preceding the 
explosion.

A magnetic
field wound up by turbulence  increases with time, until the field strength
becomes strong enough to resist the turbulent flow. Since we are interested
in weak fields, we may ignore here such backreaction by the Lorentz
forces. The precise rate of increase of the field depends on the details of
the small--scale flow. For a flow dominated by a single length scale Kraichnan
(1976) has derived exponential growth on a time scale of the order of the
turnover time $\tau$ of the flow. Thus, 
wrapping up of field lines by a small
scale flow can enhance the intensity of the magnetic field exponentially:

$$ B_{\rm seed} \ e^{t/ \tau} {\rm where} \ \ \ \ \tau \approx l/v_{\rm
turb}\eqno(10)$$

\noindent
$\tau$ being the typical turnover time of the convective cells, which is of
the order of the ratio of characteristic length of the turbulent region over
the turbulent velocity. If the duration of a turbulent period allows for a
few turnovers of the turbulent material  ($t$ larger than $\tau$),  the lines
of the magnetic field would be wrapped a few times, and  the intensity would
increase. This effect is not the classical dynamo where the field
is amplified as cyclonic motions twist the lines of the magnetic field in a
rotating fluid. The action examined here is linked to the turbulence--induced
winding up of the lines as the eddies carrying the seed magnetic field undergo
several turnovers.

{\it Therefore, prior to explosion, during the
quasistatic burning of C}, an enhancement of $B$ can occur. 
The result depends on
the duration of the turbulent quasistatic phase.

If the turbulence  lasts for more than a few turnovers,
 equipartition of the kinetic energy
density and magnetic field density could be achieved. In equipartition:

$$ {1 \over 2} \rho v^{2}= {B^{2} \over 8 \pi}\eqno(11)$$

\noindent
Taking characteristic values for the turbulent kinetic energy density,
i.e 10$^{11}$  erg  g$^{-1}$ and given that $\rho$ is $\approx$
 2--3 \ 10$^{9}$ g
cm$^{-3}$, $B$ values for equipartition are $\approx$ 10$^{10-11}$ G.

The convective phase prior to explosion is found 
 in the work by Woosley (1990) 
to last for $\approx$ 10$^{2-3}$ s. Turbulent velocities are   
of the order of a 10$^{5-6}$ cm $^{-1}$ s $^{-1}$ and the
convective core 
is a fraction of the WD radius  of typical l $\le$ 10$^{7}$ cm. This
implies turnover times of  $\tau$ $\approx$ 100 s.
In the work by Arnett (1996) and Bravo et al. (1996), the pre--explosion
evolution of the WD during the accretion process is followed in detail 
several thousand years prior to explosion.  Both works find independently
that a convective core develops several thousand years prior 
to the explosion. Turnover timescales are of the order of 300s. The 
convective period is of the order of 10$^{10}$ s, long enough, according 
to (10), to rise the magnetic field strength to equipartition values. 
Such a convective core is linked to the evolution towards explosion of 
centrally ignited Chandrasekhar C+O WDs. In edge--lit detonations
of sub--Chandrasekhar WDs, there is no such development of convection
in the C+O core in the presupernova evolution according to the 
calculations by Nomoto (1982) and Hernanz et al. (1997), since 
in  this kind of explosion there is no strongly peaked central
heating of the WD before the shock wave generated at the surface reaches
 the center and induces the explosion. 
Therefore, the initial field configuration in the core of the WD
 would not have been 
significantly distorted. The post--explosion configuration for the two
types of explosion  would then be
different and its 
effect on the light curves can help to determine the SNIa mechanism.

Thus, depending on the evolution towards explosion,  
the dynamo might have had enough time to efficiently increase
the intensity of magnetic field by large factors, or fail to do so.  
The effects of a failure to increase drastically  the mean
intensity of the field  will be reflected by the supernova light curve.

If this phase fails to develop an entagled field, the following phases 
do not favor any major change.

{\it When the incineration starts}, the turbulent velocities increase to 10
$^{7}$ cm s$^{-1}$, and  the characteristic size of the turbulent region
is of the order of 10$^{7}$ cm (Niemeyer \& Hillebrandt 1997). However, 
this phase lasts only $\approx$ 1
sec, and it would not be able to provide a sufficient enhancement of the
magnetic field in the ejecta.

After the explosion of the WD, the ejecta undergo a large expansion. The
homologous expansion achieved about 1 sec after the explosion suggests the
further conservation of the magnetic flux (there is no compression which
would distort the number of  lines crossing a given element of area). In a
homologous expansion all components of the field decrease like:

$${ B_{1} \over B_{0}} = {(R_{0})^{2} \over (R_{1})^{2} } = {(R_{0})^{2}
\over (v \times t )^{2}}\eqno(12)$$

Though the overall flow is nearly homologous on a large scale, there is 
likely to
be some form of small scale motion inside the ejecta. For example, this could
be a remnant of the convective motions in the pre-SN stage, or the result of
a Rayleigh--Taylor instability at an early stage during the explosion. Such
motions wind up field lines, causing again a
 roughly exponential growth of the
field strength, in the kinematic (low field strength) limit. The question
thus arises if such a process could increase the field strength over that
expected from a purely homologous expansion. We can now show that this is not
the case except in the unlikely event that the overturn time of the small
scale motions is less than the expansion time scale.

Small scale motions generated at any time $t_0$ during the expansion expand
with the flow, so that their length scale varies as $t/t_0$. The flow
velocity in these motions will remain the same or decrease in the presence of
dissipation, hence the turnover time scales at least as $t/t_0$.
 Hence in the absence of overall expansion,
we would expect the field to increase as ${\rm d}\ln B/{\rm d}t\approx
1/\tau$. Due to the expansion the overturning time varies as $\tau=\tau_0
t/t_0$, where $\tau_0$ is the initial overturning time and $t_0=R_0/v$ the
initial expansion time scale. Thus, because the overturning time increases
with time, the growth of the field is no longer exponential. The expansion
(using eq 12) changes the field at a rate ${\rm d}\ln B/{\rm d}t\approx
-2/t$. Adding these contributions, we get

 $$ {{\rm d}\ln
B\over{\rm d}t}={1\over\tau}-{2\over t}={1\over t}
\left({t_0\over\tau_0}-2\right)\eqno(13)$$

This shows that the decrease of the magnetic field by expansion dominates 
as long as the initial turnover time of the small scale motions is 
sufficiently long compared with the initial expansion time scale. For an 
initial expansion time scale of less than a second, this condition is easily 
satisfied by any small scale motions in the pre--SN. We conclude that the 
first term on the right in (13) can be ignored, and that the magnetic 
field decreases according to eq (12), even in the presence of small scale 
motions in the expansion.

Whatever magnetic field was present before the onset, at the quasistatic
burning phase of C, will be directly left to the effects of the overall
expansion of the ejecta, which tends to lower its intensity.

{\it After the explosion}, the flux of the magnetic field would be preserved,
and should be decreasing  with t$^{-2}$ as the supernova expands. 
{\it At 100 days the intensity of the magnetic field would have decreased by
a factor of $10^{-15}$}.  At 1000 days it would be decreased by 10$^{-17}$.
The ejecta can thus host magnetic fields much lower than the intensity of the
interstellar magnetic field, if the magnetic field in the WD has not been 
significantly enhanced prior to explosion. 
 Taking, for instance, an initial magnetic field of
10$^{4}$ G, at 100 days it will be as weak as $10^{-11}$ G, and at 1000 days
it will be $10^{-13}$ G.
The magnetic energy density will be among the lowest ones in
known astrophysical objects. The supernova becomes a huge de--magnetized
bubble (with a material density much higher than the surrounding medium,
though).

The low values of B should, however, be compared with the
dimensions which the ejecta have achieved.

For a charged particle of charge $q$, moving in a magnetic field, B, with
velocity $v$ whose component perpendicular to the direction of the magnetic
field is $v_{trans}$, the gyroradius (or Larmor radius) is:

$$r_{\rm gy} = {mc\gamma v_{\rm trans}\over qB}\eqno(14)$$

\noindent
where $m$ is the mass of the particle and $\gamma = [1 -
(v^{2}/c^{2})]^{-1/2}$.

For a positron of energy 1 MeV,  the gyroradius would be:
     
$$r_{gy} = { 4.7 \times 10^{3} {\rm} \over B}\eqno(15)$$

Thus, the gyroradii of the e$^{+}$, as compared with the size of the
envelope would be:
    
$$ {x_{\xi}} = {r_{gy} \over R} =  { 4.7 \times 10^{3}{\rm cm} \ B_{0}^{-1}
R_{0}^{-2} \ ( v \ t)}\eqno(16)$$

This is of the order of:

$${x_{\xi}} = 10^{-5} \ \ t_{7} \ \ B_{08}^{-1}\eqno(17)$$

\noindent 
where $B_{08}$ is the magnetic field before expansion in units of 10$^{8}$ G
and $t_{7}$, the elapsed time since the explosion in 10$^{7}$ s. 
 Depending on the 
$B_{0}$ value, the positrons can be more or less trapped inside the ejecta.
Less energetic particles have smaller gyroradius.  When the magnetic field 
 has
diluted down to very low  values, the gyroradius  encompasses a high fraction
of the expanded ejecta.

A pre--expansion  magnetic field of 10$^{11}$ G would prevent large
departures from full--trapping of $^{56}$Co at phases even later than 1000
days after explosion. The confinement requirements are 
seen from equation (17). The available observations on the late bolometric
decline of the light curve allow us to estimate if  a departure from the
full-trapping decay line occurs in SN. {\it This leads to an estimate of the
intensity of the magnetic field achieved before the homologous
expansion.}

\bigskip

\noindent{\it 4.1. Possible magnetic shield around the supernova ejecta}

\bigskip

 The field strength in the ejecta 
 may become 
sufficiently low after 100--1000d to allow the positrons to travel 
through the ejecta without interacting with the field.
 If the magnetic field of the WD is a dipole
 magnetic field, and the ISM around the ejecta is dense enough to 
 present significant opposition 
 to the magnetic field expansion, the conditions on final escape
 of positrons might change. 

\smallskip   

 Before the particles can be regarded 
as having 
successfully escaped from the ejecta, it must be shown that 
they are not 
reflected back into the ejecta by an external medium of sufficient density. 
 This external medium 
consists 
of two regions. Immediately outside the ejecta is a magnetic `shell', a region 
dominated by the external magnetic field of the original pre-SN core, 
now expanded 
but still containing the original amount of magnetic flux. Outside this 
shell is the 
ISM, modified by the SN shock that has passed through it. The magnetic 
pressure in 
the shell has to balance the ram pressure of the ISM relative 
to the ejecta. This 
yields its field strength:

$$ B_{\rm s}=v(4\pi\rho_{\rm I})^{1/2}=0.04 n_{\rm I}^{1/2}v_9 {\rm G}\eqno
{(18)}$$
where $v_9$ is defined such that 
 $v=10^9v_9$ is the velocity of the ejecta and $n_{\rm I}$ the
 ISM particle 
density. The flux of field lines crossing the magnetic equator 
is conserved during 
the expansion, and is of the order $2\pi B_0 R_0^2$, where  $R_0$ 
and $B_0$ are the 
radius and surface field strength of the WD.
 At time $t$, the 
thickness $d$ of the 
shell is therefore: 
$$ 
d=R_0^2B_0/(vtB_{\rm s})=2\,10^7 {B_{04} R_9^2\over
 n_{\rm I}^{1/2}v_9^2t_7}\quad{\rm 
cm} \eqno{(19)}$$
where $B_{04}$ is the
magnetic field in units of 10$^{4}$ Gauss, $R_{\rm 9}$ is the radius of
the WD in 10$^{9}$ cm.
The gyroradius of a 1MeV positron in this field is:
$$ {r_{\rm L}\over d}=5\,10^{-3} B_{04}^{-1}v_9t_7/R^2_9 \eqno{(20)}$$
Since the field is parallel to the interface with 
the ejecta, the shell presents an 
effective `shield' which reflects the positrons. 

\bigskip

Thus we may reasonable assume the shell around the ejecta to be a near 
perfect reflector. If the positrons inside it are unconfined
due to the absence of a tangled field, they will
spread uniformly through the ejecta.  The mass of the ejecta is 
concentred towards the innermost radii. If the half--mass radius is at a 
fraction $f\sim 0.1-0.2$ of the radius of the envelope, the central density
of the uniformly spread positrons is of the order $\sim 2f^3$ times 
what it would be if the positrons stayed trapped near their source in the
high--density regions. Since the luminosity is proportional to the 
density of positrons in the region containing most of the mass, the
reduction of the luminosity is likely to be a large factor, even if none of
the positrons actually escape from the ejecta.

\bigskip

\section{Bolometric declines and their interpretation}

\subsection{Physical conditions in the SN envelope and their effects
on the departures in the bolometric light curve}

  Both a weak magnetic field and the progressive thinning out of the ejecta
  produce a departure in the bolometric light curve of supernovae from 
  the full--trapping of the $^{56}$Co--decay energy. 
  There is no possibility of having 100 $\%$ trapping of positron energy
  as time goes by. The case where this departure occurs at the earliest, is
  when there is no confinement of positrons at their site of origin, as
  discussed before.  
  If the positrons are freely streaming or escaping 
  through a radial magnetic field, the time 
  required by the relativistic  e$^{+}$
  to cross the ejecta is shorter than any relevant timescale for modifications 
  in the physical conditions of the envelope, such as the
  the radioactive timescale  for $^{56}Co$--decay (111.26 days) 
  or the expansion timescale ($n / {\dot n} = t/3$  days), both being 
  of the order of 100 days. 
  In the free--streaming condition, 
  the moment when the departure occurs is early enough to ensure that any 
  deposited energy is radiated in a short timescale through collisional
  excitation and emission in a large number of forbidden transitions of 
  iron ions. Freeze--out conditions for re--radiation of that energy occur
  much later.

\smallskip

 In the confinement phase, the
  positrons do not move from their site of origin. There is a time when the    
  probability of interaction with the ions through impact ionization and
  excitation becomes very low. Although the follow--up of the
  deposition of energy by the energetic particles involves to keep 
  track of the old positrons during the whole expansion history of the envelope 
  while injecting the new ones at each site, the fact of neglecting them 
  when they start to become very 
  inefficient, i.e  after $ t > t_{c}$ (see section 2),  
  is a fair approximation. The
  probability for interactions decreases with t$^{-3}$ and the contribution of
  all those positrons in the diluted medium is much smaller than before, at
  $ t > t_{c}$. At the latest times, 
  the  quantitative prediction can underestimate somehow the luminosity. 
  The estimate of the time at which the departure
  occurs according to the physical SN model and magnetic field configuration,
  is, however, very precise. 
 
\bigskip 
 
\noindent{\it Timescales  and expected departures}

\noindent
 The frequency of impact ionizations or excitations by the 
 $e^{+}$
 becomes lower as $n_{e}^{+}$, the density of 
 the energetic positrons, and that of the target ions decrease.

\noindent
 The timescale for impact ionizations by positrons, $\tau_{e^{+} coll}$,
 can be expressed as: 

$$\tau_{e^{+} coll} = \left[\int_{E_{min}}^{E_{max}} n_{e^{+}} (\tilde E)
\int_{0}^{\tilde E} \sum_{ij} \sigma_{ij} (E)\ v_{\!E}\ f_{ij}\
dE\right]^{-1}\eqno(21)$$

\noindent
where  $n_{e^{+}} (\tilde E) $ is the number density of positrons 
of a given energy $\tilde E$ originated in the $^{56}$Co decay, 
 $\sigma_{ij}$ are the impact ionization cross sections with each 
 target ion $\it i$ of species $\it j$ and $f_{ij}$ are the relative 
 abundances of those ions. $\rm E_{min}$ and $\rm E_{max}$ are the 
 minimum and maximum kinetic energy of the positrons, and $\rm v_{\!E}$
 the velocity. $n_{e^{+}} (\tilde E) $ is much lower than $n_{e}$, 
 the electron density.  Such timescale increases due to the 
 thinning out of the ejecta, and at a given point it becomes 
 larger than the radioactive 
 timescale  and the expansion timescale.

 Equivalently, the timescale for the positrons
 to lose half of their energy, $\tau_{e^{+} loss}$, becomes
 also large (this quantity is related to the stopping distance 
 of the positron, which grows with time, see Table 1): 

   $${\tau_{e^{+} loss}} \sim  { E  \over   \dot E}  \eqno(22) $$

\noindent
  In {\it the confinement regime}, the rate at which interactions occur 
 is favored by the presence of a uniform density of target ions
 along the positron path. The departure occurs much later than in
 {\it the free--streaming regime} for all models. 
  This departure signals a point of ``breakout
 of nonthermal ionization balance'' or  ``non--steady state for the 
 nonthermal processes''. However, the collisional processes and radiative
 transitions between levels still occur at a fast rate.
 Reemission is occuring through the large number of forbidden transitions
 of iron ions. The deposited energy is reemitted, to a very 
 high degree, in steady state. 

\bigskip
 
\noindent
 In the {\it positrons free streaming regime}, 
 $\tau_{e^{+} loss}$ for the most energetic
 positrons is longer than 
 the crossing time of the envelope. 
 Those positrons, which are not confined, 
 will escape with high kinetic energies. The free streaming favors much
 earlier departures from full--trapping of the $^{56}Co$, at epochs when
 the recombination and collisional processes still occur at high rates.

\bigskip

\noindent
{\it The infrared catastrophe: freeze--out of the supernova ejecta}

\noindent
 The observational requirements to 
 extract information from the bolometric light curve of a supernova 
 for the prospects given in this work are
 to rigurously reconstruct this light curve placing estimates
 on the infrared emission, and to  complement this task with a spectrum
 at the time where a change in the decline rate is observed. 
 Here, 
 we present 
 our predictions to be compared with future observational data. 
 By obtaining those data, it should be possible to 
 distinguish observationally between the various effects entering 
 in the bolometric light curve decline.

\bigskip

\noindent
 Along this work, when addresing the departure in the luminosity, we 
 assume that the emission at ultraviolet, optical and infrared
 broad bands is recovered, and 
 an estimate of the temperature or evaluation of  
 how much luminosity has gone into far infrared wavelengths has been done.

\bigskip

\noindent
  Axelrod (1980) first pointed out that the supernova ejecta in their
  late--time evolution would reach a point when temperatures would 
  fall below a critical temperature,
  T$_{c}$ $\simeq$ 2000 K, in their innermost layers. When this occurs,
  most of the emission of the supernova, would come out in the
  fine structure forbidden transitions at infrared wavelengths. 
  This is named as the Infrared Catastrophe (IRC) since it will imply an
  inflation of the emission at very long wavelengths
  while a depletion in the optical and ultraviolet emission occurs. 
  It is possible to calculate for each model when $T_{\it core}$ (in the 
  innermost dense ejecta)   
  falls below T$_{c}$, and determine it as well observationally.
   
\bigskip

\noindent
  The departure, when an IRC occurs, affects the B, V, R monochromatic light 
  curves, but it is a temperature effect and does not imply a proper 
  departure of the overall emissivity. 
\bigskip

 Limitations in the accuracy of the results of the following sections
 arise from the uncertainties 
 in the observations and
 reconstructions of those bolometric light curves (if a limited number of
 photometric data are available) and from the growing time--dependence
 of the reemission processes well after two years.    
 The different predictions related to the two evolutionary 
 histories of centrally ignited Chandrasekhar WDs and edge--lit 
low mass C+O WDs seem worth testing.   
  Combining the analysis of the
 $\gamma$--ray tail and the positron tail helps to complement information 
 on the physical models. This information will be addressed in the next 
 sections.

\subsection {Physical models and the rate of the late decline of SNe Ia}

 As a general
trend,  C+O centrally ignited Chandrasekhar WDs tend to trap
significantly the positrons and give a bolometric light curve decline close to
the full trapping line drawn by the exponential decay of $^{56}$Co. The
bolometric light curves of sub--Chandrasekhar models tend to fall below the
full--trapping line after 400 days even if $B$ confines the e$^{+}$, or even
earlier in massive edge--lit detonations (model NIDD by Nomoto 1995). A
follow--up of those bolometric light curves is a good tool to clarify the
nature of the explosion. The bolometric light curve in the earlier $\gamma$--ray
dominated tail (before 200--300 days) is different for those models and
allows a first discrimination. The positron tail informs further about the
ejected mass and the magnetic field configuration. 
In Figure 4 we present M$_{bol}$ decline rates 
for different models during the first 400 days under the
confinement hypothesis. This sort of figure can be useful for 
comparison with observations. 
  Departures from the full--trapping of
$^{56}$Co--decay  of the order of 10--15$\%$ at about 400 days can be explained
by the distribution of radioactive material. Larger departures, of 
30--40 $\%$ or
larger,  have to be interpreted in terms of lack of magnetic field
confinement of the positrons, or even enhancement of the escape.

\subsection{Trapping and departure: the magnetic field and the 
 mass of the ejecta}

A way to evaluate from its physical effects the real effectivity of positron
trapping is to compare the calculations with observed bolometric light curves
of SNe Ia. Very few bolometric light curves of supernovae are, however, 
available. For SN 1972E in  NGC 5253, about two years of  bolometric
follow--up after the explosion is provided by Kirshner \& Oke (1975).  
Since this exceptional long--lasting coverage, the general trend has
 been to concentrate 
the observations of supernovae  to the first year after
the explosion. Suntzeff (1996) obtained the 
bolometric light curve of SN 1992A 
 up to 300 days after explosion. More recently, the 
fast declining bolometric light curve of SN
1991bg covering the first two hundred days after explosion  was 
presented by Turatto et al. (1995).

What can we learn from the observations? SN 1972E  might represent a SN Ia
close to ``normality'' in its luminosity and spectral characteristics (i.e.
similar to SN 1981B, SN 1990N). The spectral scans obtained by Kirshner \&
Oke (1975) and integrated along wavelength by Axelrod (1980) provide a
bolometric (or quasi--bolometric) light curve which can usefully be compared 
with model calculations. 
The distance to the supernova is known from the Cepheids period--luminosity
relationship (Sandage et al. 1994), and it is known that the supernova
was not substantially reddened ($ E(B-V) \approx 0.05$). We scale the
absolute luminosity values according to the known distance 
to NGC 5253 and a low E(B--V) (Axelrod had assumed E(B--V) =0.22), 
and compare them with the model predictions. Figure 3 shows that the 
SN 1972E bolometric light curve follows
well  the behavior of centrally ignited C+O Chandrasekhar WD with a
turbulent magnetic field configuration. In particular, model W7 seems 
to be giving a very good
account of the bolometric light curve.  The turbulent configuration of the
magnetic field is thus favored by the level of deposition of energy  suggested
by  the bolometric curve of SN 1972E prior to 500 days. A lack of  magnetic
field as well as a radial strong magnetic field would produce larger
departures from the full--trapping $^{56}$Co--decay line.

The model light curve follows well the observed one until at least 500 days.
Then, a departure at 720 days after maximum is observed, of the order of  70
$\%$ enhancement of escape as compared with the confinement prediction (only 30
$\%$ of $^{56}$Co positron kinetic 
energy is deposited). The confinement prediction gives
only 10 $\%$ of departure at 740 days for model W7. Taken at face value, 
this reported departure, and the 
elapsed time in the
confinement regime (in the case of SN 1972E up to 500 days according to our
analysis), tell us  
about the magnetic field intensity prior to explosion, as
discussed above.  According to equation (17), the confinement of positrons 
starts
to fail at about 700 days for WD magnetic fields of $B \approx 10^{5} G$, a
plausible value. If the magnetic field intensity of the initial WD
would have been lower, the departure would have occured earlier. A very 
magnetized WD prior
to explosion (B $\ge$ 10$^{10-11}$ G) would not give any
significant departure until much
later on. Unfortunately, it can not be discarded that the very 
last point in the light curve is more unaccurate
than the rest of the data (Kirshner 1997), 
and that the bolometric light curve keeps 
falling not far from full--trapping. If that were the case, and the bolometric
light curve would follow within a 10$\%$ of departure the 
full--trapping curve, that would
confirm the theoretical expectations of a chaotic magnetic field enlarged
up to equipartition during the long convective accreting period expected 
in centrally ignited C+O WDs. The safest conclusion given here is that
{\it SN 1972E represents the case of a centrally ignited C+O WD with a likely
tangled magnetic field of at least 10$^{5}$G prior to the expansion}. The 
light curve and luminosity is well reproduced by model W7.

Future observations, describing the bolometric light curves of SNe Ia, should 
provide information about the post--explosion magnetic field. Strong
dust obscuration can cause deviation in the bolometric declines, but it is
accompanied by shifts in the centroids of the emission lines at late phases.
Thus, there is a way to point out when dust obscuration occurs. The light
curve can be corrected by observing the far infrared and including the energy
emitted at those wavelengths. The bolometric light curve of SN 1992A seems
to follow the same decline rate as SN 1972E, although data are only
available up to 300 days. Given the uncertainties in the distance to this
supernova, we have shifted arbitrarily in the figures
 the absolute scale of the luminosity
given by Suntzeff (1996).

\subsection{ A fast decaying bolometric light curve}

A much faster decline than in SN 1972E is seen in SN 1991bg. Observations
were presented in terms of the {\it uvoir} bolometric light curve 
by Turatto et al.(1995). They followed the same procedure to 
integrate the luminosity in the
different bands as Suntzeff (1996) for SN 1992A. Light from the
far--infrared is not included in their luminosity count. However, {\it JHK}
observations by Porter et al. (1992) showed a fast decline of 3 mag in the
first month and no secondary maximum. This suggested that the supernova was
not emitting strongly in the infrared.

Calculated bolometric light curves are displayed in Figure 5 and compared
with SN 1991bg. The decline after 200 days (and even earlier) is faster than
predicted for confinement by a chaotic magnetic field in a wide variety of
models. The early decline shows that the two opposite models proposed to
explain this supernova: an edge--lit detonation (Livne \& Arnett 1995) and a
pulsating delayed detonation model in a Chandrasekhar--mass WD 
(Woosley 1997) fail
to give the right luminosity and evolution in time of the bolometric light
curve in the $\gamma$--ray dominated phase already. Shifting the absolute 
magnitude
of both models to agree with the observed one would imply as well uncomfortable
distances to the core of the Virgo Cluster (well beyond the current
discussions on it). The bolometric light curve of SN 1991bg requires a 
small mass of
$^{56}$Ni, of the order of 0.07 M$\sb\odot$, as found by spectral modeling
(Ruiz--Lapuente et al. 1993). In addition to requiring a small mass of
$^{56}$Ni, further considerations are needed to explain the unusual late
behavior.  At day 200 there is a departure from the confinement prediction: 
the deposition is only 50$\%$ of what would be expected in the confinement
case. In this case the observational basis for such departure is firmly
established. Different options have to be considered: 1) a low magnetic
field of the original WD precludes confinement. To evaluate this option,
positron escape in the absence of a magnetic field is calculated for both a
Chandrasekhar and a sub--Chandrasekhar explosion model (model W7 and model
WD065 of the detonated WD of 0.65 M$\sb\odot$). 2) A radially combed--out
magnetic field enhances escape in a low--mass WD explosion and in a
Chandrasekhar WD explosion (same models and model 2 of the edge--lit
detonation of a 0.7 M$\sb\odot$ WD). None of the hypotheses combining a
Chandrasekhar mass model and enhanced escape can account for a 50$\%$ of
deposition of the $^{56}$Co--decay energy at 200 days. As shown in Fig. 5,
the option of lack of a significant magnetic field  and a small ejecta mass,
as in model WD065, gives a reasonable agreement  with the observations. In
the absence of a magnetic field, positrons lose their energy according to the
interactions undergone along their free trajectories. The sort of
calculations done here rescale the $\gamma$--ray results to an opacity
appropiate for the processes undergone by the e$^{+}$: $\kappa_{e} \sim 10\
g^{-1}\ cm^{2}$ (Axelrod 1980; Colgate, Petschek, \& Kriese 1980). The
calculation then reproduces the observed decline rates.

The picture described here of positron transport in the ejecta
corresponds to  zero confinement or negligible action of the magnetic field.
Such condition occurs when the magnetic field intensity of the WD prior to
explosion is lower than 10$^{3-4}$G. Thus, this looks like a plausible
explanation for SN 1991bg: {\it low--mass and weak magnetic field prior to
expansion. Lack of confinement but a Chandrasekhar--mass explosion gives a
light curve 
much closer to full-trapping and it departs much later than
observed}.

The option of the enhanced escape through radial magnetic fields had to be
evaluated by integrating the trajectories of the positrons. Escape occurs in
Chandrasekhar--mass explosions at a level lower than observed. Low--mass and a
radially combed but strong magnetic field is thus another possible explanation,
although less likely from the implications of evolution in time of the
magnetic field. 

To summarize, the bolometric light curve of SN 1991bg can be well accounted
for if positrons are not confined by the post--explosion magnetic field,
 due to a low initial magnetic field or
a radially combed--out magnetic field in low--density ejecta (small mass). 
 The observed
light curve seems  hard to fit with Chandrasekhar--mass WD explosions, 
since even for the most favorable configuration of the magnetic field
to enhance escape, a Chandrasekhar mass of ejecta would be enough to produce 
significant deposition
of  $\beta^{+}$ energy.

\subsection{ Bolometric light curves and the mass in Type Ibc}

The mass of the star at the time of the explosion in Type Ibc is a matter of
discussion, and it is linked to the identification of their progenitors. The
well observed bolometric light curve of SN 1994I (Richmond et al. 1995)
allows us to discuss models for Type Ibc SNe as compared with the
observations.   The precursors of Type Ibc could be Wolf--Rayet stars, with
main sequence masses in the range of 30--40 M$\sb\odot$. Those stars undergo
strong winds and also mass transfer, if they are in binary systems. Mass
transfer and winds might produce the loss of the  H envelope in the star
without removing completely the He envelope.   After undergoing gravitational
collapse at the end of their evolution,  a total ejected mass close to  2
M$\sb\odot$ is expected from this massive progenitor case (Woosley,  Langer,
\& Weaver 1996).  In other scenarios, the initial mass of the progenitor is
smaller--i.e. in the range of 10--20 M$\sb\odot$--, and the star ends up its
evolution, after having lost both the H envelope and the He mantle, as a bare
C+O core. The ejecta mass could be below 1 M$\sb\odot$.  Both $\gamma$--ray
and energy deposition by positrons should be different in the two cases.
Spectral calculations show the need of enhanced mixing in SN 1994I, and in
other SNe Ic (Eastman \& Woosley 1997; Ruiz--Lapuente 1997). Large--scale
mixing, required for  SNe Ic, will affect very much the deposition of energy
by positrons, and thus the bolometric  luminosity.  Figure 6 shows the
difference that mixing induces in the deposition of $\gamma$-rays in Type Ibc
models. Mixing enhances  as well the escape of energy from  positrons. In
ejecta with masses lower than 1 M$\sb\odot$ it leads to a departure from the
full--trapping curve of $^{56}$Co decay.  Figure 7 shows the deposition of
energy from e$^{+}$ and its evolution in time for different models of SNe Ibc
and the chaotic configuration of the magnetic field. Model 7A corresponds to
the more massive progenitor option mentioned above, and model 7A mixed has
the  same ejected mass, but with large--scale  mixing  (Woosley, Langer, \&
Weaver 1996; Eastman \& Woosley 1996), as required from the spectra. Model
CO21 (Nomoto et al. 1996) corresponds to the less massive progenitor, and a
mixed version has also been calculated.

It can be seen, that the positron energy is not deposited in the ejecta of
exploded C+O stars already at 200 days after explosion. A more massive
ejecta than 1 M$\sb\odot$ seems to be 
required to produce effective trapping of the energy
from $\gamma$--rays and positrons and preclude a fast decline of the
luminosity. As shown in Figure 8, such requirement gives a good account 
of the observed bolometric light
curve of SN 1994 I.

\section{ Positron escape and the Galactic 511 keV line}

The positron annihilation radiation towards the Galactic center (Haymes et
al. 1975) is believed to originate from two contributions: a time--variable
compact source located in the Galactic center and a diffuse component along
the Galactic plane. Supernovae have been identified as a likely origin for
the diffuse component: positrons escaping from supernovae and annihilating in
the surrounding regions would give rise to that emission (Lingenfelter \&
Ramaty 1989). The width of the diffuse 511 keV radiation places strong
constraints on the temperature and density of the region where the
annihiliation takes place (Ramaty \& M\'ez\'aros 1981; Guessoum, Ramaty \&
Lingenfelter 1991; Wallyn et al. 1993). It is found that electron--positron
pairs need to lose their energy in a dense medium before annihilating, or the
511 keV would be broadened and blueshifted (Ramaty \& M\'ez\'aros 1981). Our
findings about positron confinement suggest that the SN ejecta are a first
site where positron can be confined, as the ejecta evolve into the remnant
phase. The way in which the nonthermal positrons are retained in the
increasingly diluted ejecta until they escape to the neighbouring ISM depends
on the intensity and configuration of the WD magnetic field prior to
explosion, which is determined by the WD evolutionary path.  Type Ia
supernovae exploding as centrally ignited Chandrasekhar WDs would favor
confinement through a chaotic magnetic field, whereas  sub--Chandrasekhar
edge--lit WDs present an environment more favorable to escape of positrons
from their site of origin, although those particles find in the region of
interaction between the supernova and the interstellar medium a shield
precluding further escape.

\section{Conclusions}

This work  has shown that the bolometric light curves of 
SNe Ia trace a
poorly investigated property of the supernova progenitors: their magnetic
field. Through a well tracked departure from the full--trapping curve of
$^{56}$Co decay, insights on the pre--expansion magnetic field of the star
can be obtained. It can be investigated whether the convective turbulence
previous to the explosion in accreting WDs succeeds in amplifying the mean
intensity of the original magnetic field of the WD by winding up the magnetic
field lines. Or whether, on the contrary, the original WD magnetic field
prevails without growing significantly before the explosion and its intensity
simply decreases as the supernova expands. The consequences of these two
extreme hypothesis have appreciable different impacts on the supernova 
luminosity. It is possible that Chandrasekhar WDs develop a highly tangled
magnetic field which would favor a bolometric light curve close to 
full--trapping of $^{56}$Co decay positrons energy. The important phase
when this can occur is the period when accretion and gain in mass of the
C+O WDs lead to a compression and quasistatic C burning in the center. A
central convective core is developed several thousands years before 
the explosion. This should be a distintive signature of C+O accreting WDs
precursors of centrally ignited Chandrasekhar explosions. It is 
not expected to occur in sub--Chandrasekhar WDs ignited through He detonations.
In those explosions, the initial magnetic field of the WD would have 
preserved its initial configuration enhancing the escape of the energy
of the positrons.  

On the other hand, differences in the distribution of radioactive material in
velocity space and in the mass of the exploding WDs give rise as well to
different declines in the late--time bolometric luminosity of SNe Ia. A
tabulation of typical decline rates resulting from different explosion
mechanisms is given to facilitate comparisons with observations.  Three
different epochs can be considered in the deposition of energy from
$^{56}$Co: a first phase where $\gamma$--ray deposition is sustaining the
early $^{56}$Co tail. A second phase where the $\gamma$--ray contribution is
starting to become negligible and $\beta^{+}$--rays start to provide the
luminosity. In this second phase the density of the ejecta  (provided that
the ejected masses are close to 1.4 M$\sb\odot$) is still high enough to
ensure the effectivity of positron energy deposition. And a third phase in
which the density of the ejecta is not high enough to trap significantly the
positron energy and the configuration and intensity of the magnetic field
determines the fate of the released energy. A very wide difference in the
tails of the bolometric luminosity at this phase could easily be linked to
different intensities of the magnetic field previous to the enormous
expansion resulting from the explosion. If a wide diversity (larger than $30
\%$) is found even earlier, in the second phase --i.e, as soon as
$\gamma$-rays become a negligible energy contribution--, since at that time
the density of the ejecta should still be high enough to slow down the
positrons, the departure points towards an enhanced escape favored by the
preserved but expanded dipole structure of the original $B$ of the WD, and,
also to possible low mass of the ejecta in case of a extreme escape occuring
as early as in SN 1991bg.  A lesser degree of diversity (lower than $10-15
\%$ in deposition) in the bolometric decline at this intermediate phase could
be interpreted as differences in the kinematics and $^{56}$Co--distribution
in the supernovae.

Our analysis of SN 1972E suggests that full--trapping lasted at least
400--500 days. The Chandrasekhar model W7 accounts well for the overall shape
of the bolometric light curve. The  decline of the supernova so close to
full--trapping and the late departure taking place at about 740 days after
explosion indicates that a tangled magnetic field of at least 10$^{5}$G had
developed in this supernova. In the case of SN 1991bg, the early  and large
escape of energy suggests a lower $B_{0}$ previous to explosion (lower than
10$^{3-4}$ G),  or a dipole magnetic field expanded towards a radial
structure.   Whereas in SN 1972E there is no evidence pointing towards a mass
lower than the Chandrasekhar mass, in SN 1991bg even within magnetic field
configurations maximally favoring escape, a Chandrasekhar--mass WD would
still show larger trapping of radioactive energy than observed. A very good
agreement with the observations of SN 1991bg  is found if confinement is
negligible and the total mass of the ejecta is {\it a half of a
Chandrasekhar mass}.

On the other hand, in determining the evolution of the bolometric luminosity
of core--collapse supernovae (coming from massive stars) and their ejected
mass, mixing plays a fundamental role. A study of the mixing in each type of
core--collapse SN through spectral modeling is needed before deriving any
conclusions on the ejected mass and on the amount of $^{56}$Ni synthesized in
those explosions. The mass of the star at the time of the explosion in Type
Ibc is a subject of discussion, even in such well--observed cases as SN 1994
I. Examining the bolometric light curve of this supernova, and the mixing
constraints from the spectra, a better agreement with the more massive
progenitor is suggested. The diversity in the bolometric luminosity of Type
Ibc can be linked both to mixing and ejected mass differences among the
exploded stars giving rise to this supernova class. A longer follow--up of
their bolometric light curves will help to determine the trapping of
$^{56}$Co energy and to clarify the actual mass range of the stars which
explode as supernovae of different types.

\bigskip

P.R.L thanks the very estimulating discussions with Wolfgang Hillebrandt in
relation to this work, and useful information provided by him, Dave Arnett
and Jens Niemeyer on convection in WDs. Thanks go as well to Peter Milne for
interesting exchanges on positron transport in supernovae. Financial support
for this work has been provided by the Spanish DGICYT.

\vfill\eject

\begin{table*}
\begin{center}
\caption{ Typical values for the stopping distance and the fraction of the
envelope traveled by e$^{+}$ of various energies by ionization}
\begin{tabular}{lllll}
\hline
 & 1 keV & 10 keV & 100 keV & 1 MeV \\
\hline
W7 $d_{e} (300)$ & $5.2\times10^{11}$ & $2.3\times10^{13}$
 & $1.2\times10^{15}$ &
$2.8\times10^{16}$ \\
 \ $\xi (100)$ & $1.0\times10^{-6}$ & $4.4 \times 10^{-5}$ & $2.2 \times
10^{-3}$ & 0.05 \\
 \ $\xi (300)$ & $9.1\times10^{-6}$ & $4.4 \times 10^{-4}$ & 0.02 &0.48 \\

LA95M2 $d_{e} (300)$ & $1.1\times10^{12}$ & $4.9\times10^{13}$
 & $2.5\times10^{15}$ &
$5.9\times10^{16}$ \\
 \ $\xi (100)$ & $1.6\times10^{-6}$ & $6.9 \times 10^{-5}$ & $3.3 \times
10^{-3}$ & 0.08  \\
 \ $\xi (300)$ & $1.4\times10^{-5}$ & $6.2\times 10^{-4}$ & 0.03&0.75 \\
\hline
\end{tabular}
\end{center}

\hskip 2.25 true cm d$_{e}$: stopping distance (in cm)

\hskip 2.25 true cm $\xi$ = $d_{e}/R_{env}$

\end{table*}

\begin{table*}
\begin{center}
\caption{ Models for Type Ia $^{1}$}
\begin{tabular}{lllllll}
\hline
 & W7 & WPD1 & LA95M2 & LA95M6 & WD065 & NIDD   \\
\hline
$Mass \ (M_{\odot}) \ $ & 1.38  & 1.38 & 0.7  & 0.96 & 0.65 & 1.07   \\
$ Mass \ ^{56}Ni \ (M_{\odot}) \ $ & 0.63 & 0.1 & 0.14 & 0.65 & 0.07 &
0.97    \\
$ E_{kin} \ \ (10^{51}erg)$ \ & 1.3 & 1.1 & 0.66 & 1.33 & 
 0.57 & 1.33 \\  
\hline
\end{tabular}
\end{center}
$^{1}$Models are: deflagration model W7 of Nomoto, Thielemann \& Yokoi
(1984); pulsating
delayed detonation model WPD1 of Woosley (1997); He--detonation of a
0.7 M$\sb\odot$ WD (model 2 by Livne \& Arnett 1995) and of a 0.96
M$\sb\odot$ WD (model 6 by Livne \& Arnett 1995), a He--detonation of a
1.1 M$\sb\odot$ WD (Nomoto 1995, here called NIDD), and a bare C+O
detonation of a 0.65 C+O WD (model WD065 by Ruiz--Lapuente et al. 1993).

\end{table*}

\vfill\eject

\begin{table*}
\begin{center}
\caption{Declines in absolute magnitude in the e$^{+}$ confinement regime}
\begin{tabular}{llllll}
\hline
 & $\Delta_{m_{\gamma}}^{100}$ & $M_{bol}^{100}$ & $\Delta_{m_{1 \beta}}$ &
 $\Delta_{m_{2 \beta}}$ & $M_{bol}^{200}$ \\
\hline
 W7   & 0.021  & -16.   & 0.013  & 0.010  & -13.9   \\
 WPD1 & 0.022  & -14.3   & 0.013   & 0.010  & -12.11   \\
 LA95M2 & 0.009 & -13.6  & 0.012   & 0.010  & -11.8   \\
 LA95M6  & 0.009  & -15.1  & 0.012   & 0.011 & -13.4 \\
 WD065    &  0.017  & -12.4   &  0.012   &  0.011  &  -10.7 \\
 NIDD & 0.0125  & -14.6   &  0.011  &  0.011  &  -13.4 \\
\hline
\end{tabular}
\end{center}
The models are described in Table 2. 
\end{table*}

\begin{table*}
\begin{center}
\caption{Deposition of e$^{+}$ for different models$^{1}$ in the
confinement regime}
\begin{tabular}{lllllll}
\hline
  & W7 & WPD1 & LA95M2 & LA95M6 & WD065 & NIDD   \\
\hline
$D_{\beta} \ (300 d)$ & 100.  & 95.0 & 97.8 & 89.6 & 100 & 84.7 \\
$D_{\beta} \ (350 d)$ &  100.  & 95.0 & 95.8 & 87.2 & 100 & 78.6 \\
\hline
\end{tabular}
\end{center}
The models are described in Table 2.
\end{table*}

\vfill\eject

\begin{figure*}
\centerline{\epsfysize15cm \epsfbox{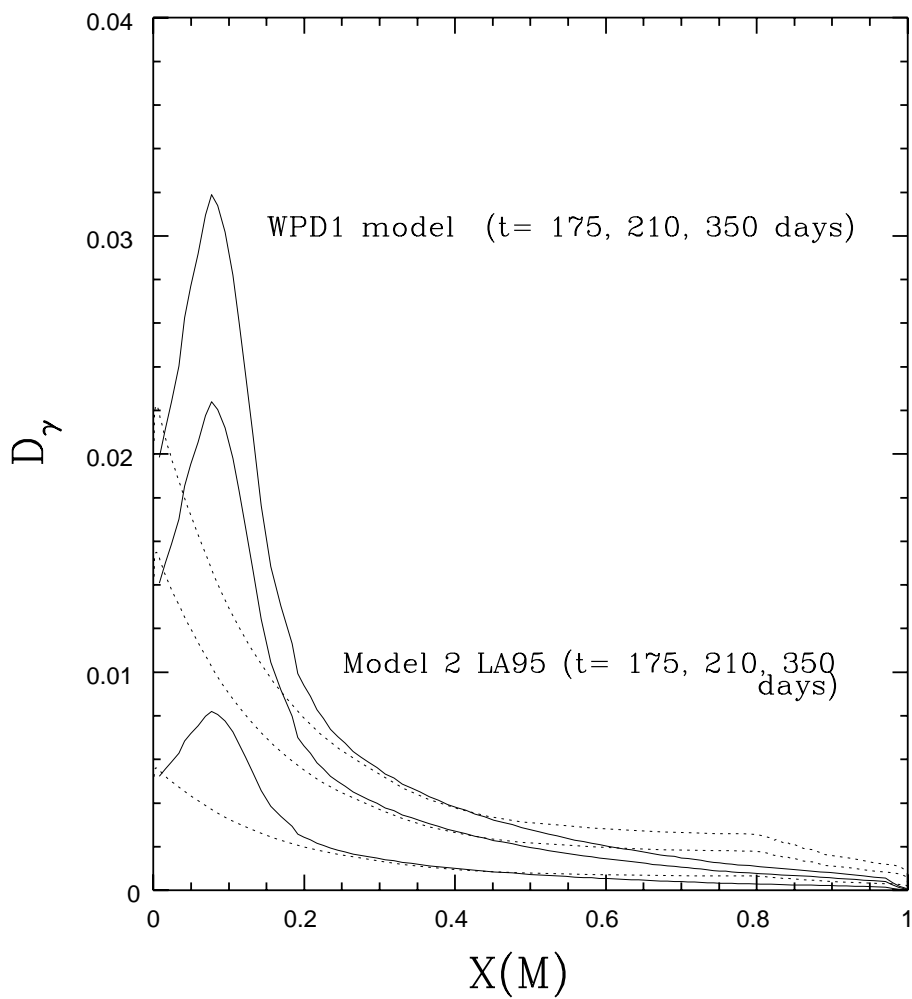}}
\nopagebreak[4]
\vspace{4mm}
\caption{Deposition of $\gamma$--rays for two alternative models for
subluminous SNe Ia: the dashed line displays
the deposition function of 
a sub-Chandrasekhar edge--lit detonation of a 0.7 M$\sb
\odot$ C+O WD (model 2 by Livne \& Arnett 1995), and 
the solid lines draws the same function for a pulsating delayed
detonation model in a Chandrasekhar C+O WD (model WPD1 by Woosley). 
The deposition profile is calculated at different times after the
explosion.}

\end{figure*}

\bigskip

\begin{figure}[hbtp]
\centerline{\epsfysize17cm \epsfbox{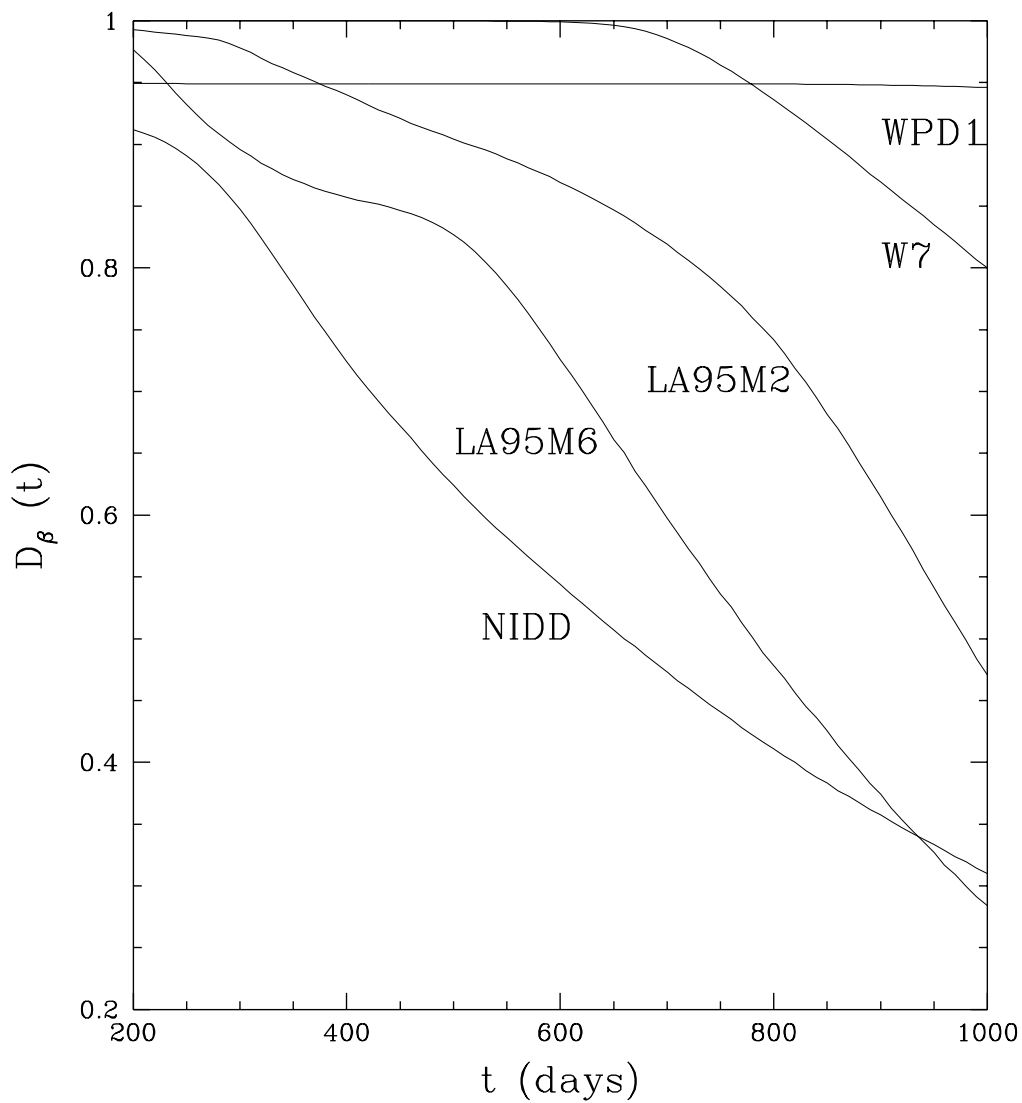}}
\nopagebreak[4]
\vspace{4mm}
\caption{Deposition of energy from e$^{+}$ for various SNe Ia models.
Models are as in Table 1: deflagration model W7 of Nomoto, 
Thielemann \& Yokoi(1984); pulsating
delayed detonation model WPD1 of Woosley (1997); He--detonation of a
0.7 M$\sb\odot$ WD (model 2 by Livne \& Arnett 1995, i.e LA95M2) and of a 0.96
M$\sb\odot$ WD (model 6 by Livne \& Arnett 1995, i.e LA95M6),
 a He--detonation of a 1.1 M$\sb\odot$ WD (Nomoto 1995, here called NIDD). }
\end{figure}

\bigskip

\begin{figure}[hbtp]
\centerline{\epsfysize12cm\epsfbox{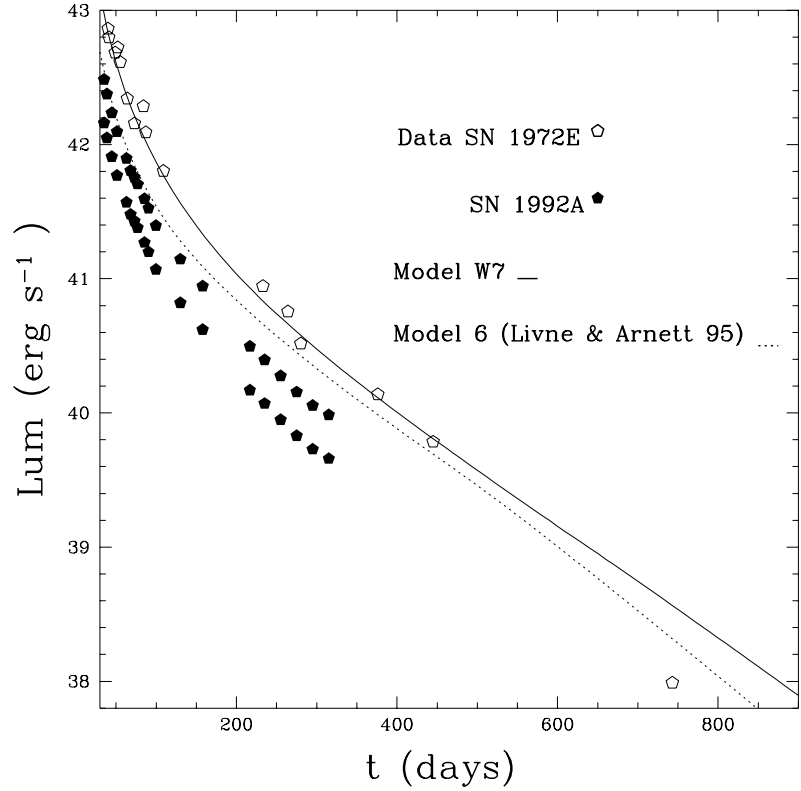}}
\caption{Bolometric light curves for the various models
 compared with the data for SN 1972E (Kirsnher \& Oke 1975)
 and the bolometric data for SN 1992A (Suntzeff 1996)
shifted in scale of distance (d=16.5 and d=22 Mpc for NGC 1380).
Model W7 by Nomoto, 
Thielemann \& Yokoi(1984) is a deflagration of a Chandrasekhar--mass
WD. Model 6 by Livne \& Arnett (1995) is a He--detonation of a 0.96
M$\sb\odot$ WD.}
\end{figure}

\begin{figure}[hbtp]
\centerline{\epsfysize13cm\epsfbox{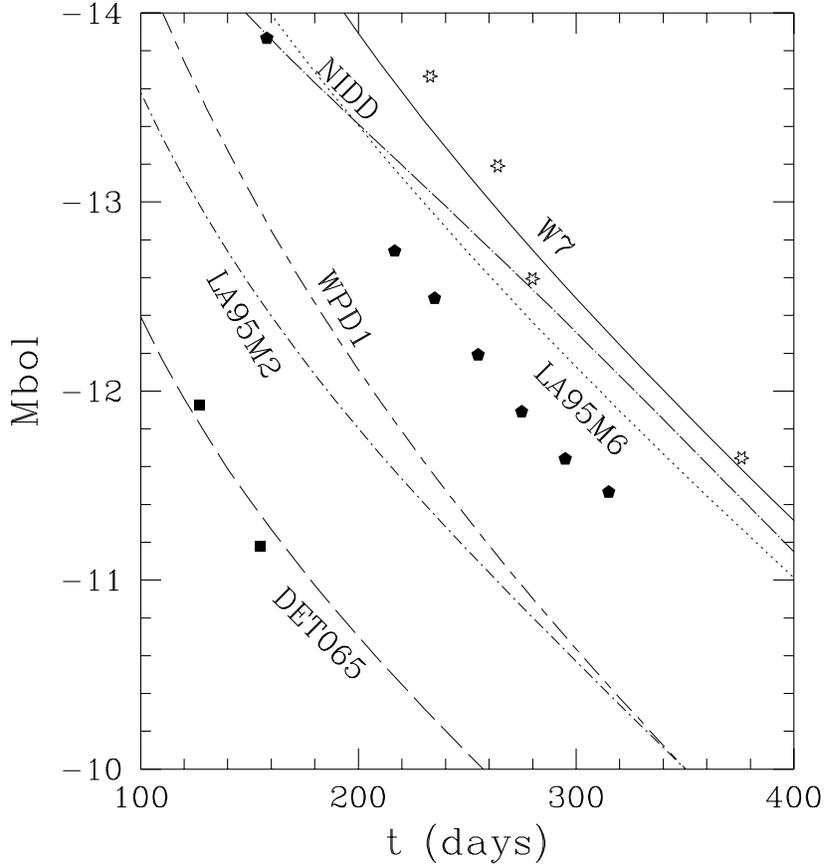}}
\caption{Bolometric magnitude--decline for models. 
 Data for SN 1972E (starred symbols), SN 1992A (pentagonal
 symbols) and SN 1991bg (square symbols) are shown for comparison. The
 models predictions are displayed by different labeled lines.   
Model W7 by Nomoto, 
Thielemann \& Yokoi(1984) is a deflagration of a Chandrasekhar--mass
WD. Model WPD1 by Woosley (1997) is a pulsatimg delayed detonation 
of a Chandrasekhar WD. The model labeled LA95M6 is 
model 6 by Livne \& Arnett (1995), a He--detonation of a 0.96
M$\sb\odot$ WD. The model labeled LA95M2 is model 2 by Livne \& Arnett (1995),
a He--detonation of a 0.7 M$\sb\odot$ WD. The model labeled NIDD is a 
He--detonation of a 1.1 M$\sb\odot$ WD as calculated by Nomoto (1995), and
model DET065 is a bare detonation of a C+O WD of 0.65 M$\sb\odot$
(Ruiz--Lapuente et al. 1993). See details in Table 2.}
\end{figure}

\begin{figure}[hbtp]
\centerline{\epsfysize20cm\epsfbox{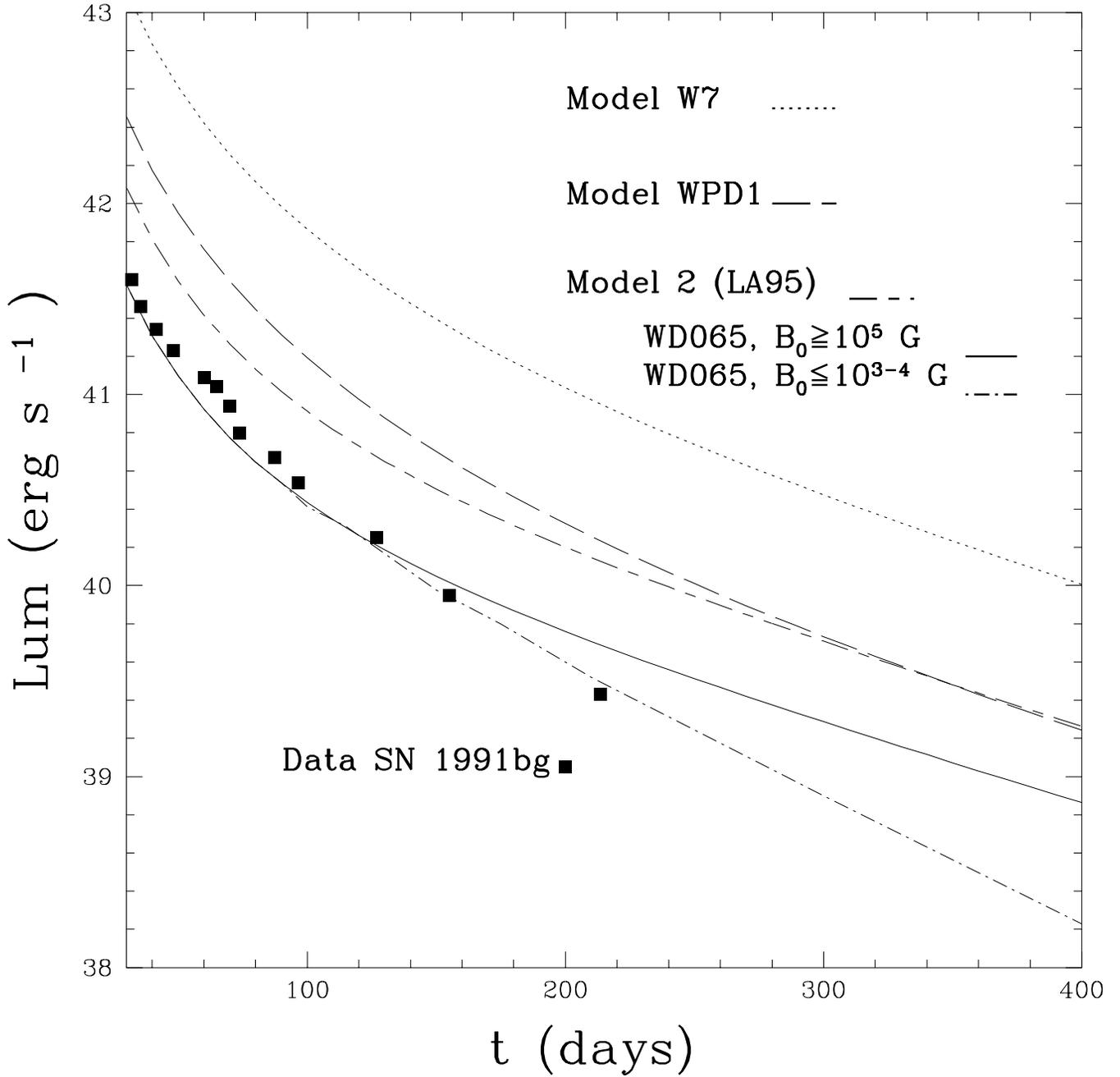}}
\nopagebreak[4]
\vspace{4mm}
\caption{Bolometric light curves for various models proposed for 
subluminous SNe Ia compared with SN 1991bg (Turatto et al. 1995).}
\end{figure}

\begin{figure}[hbtp]
\centerline{\epsfysize17cm \epsfbox{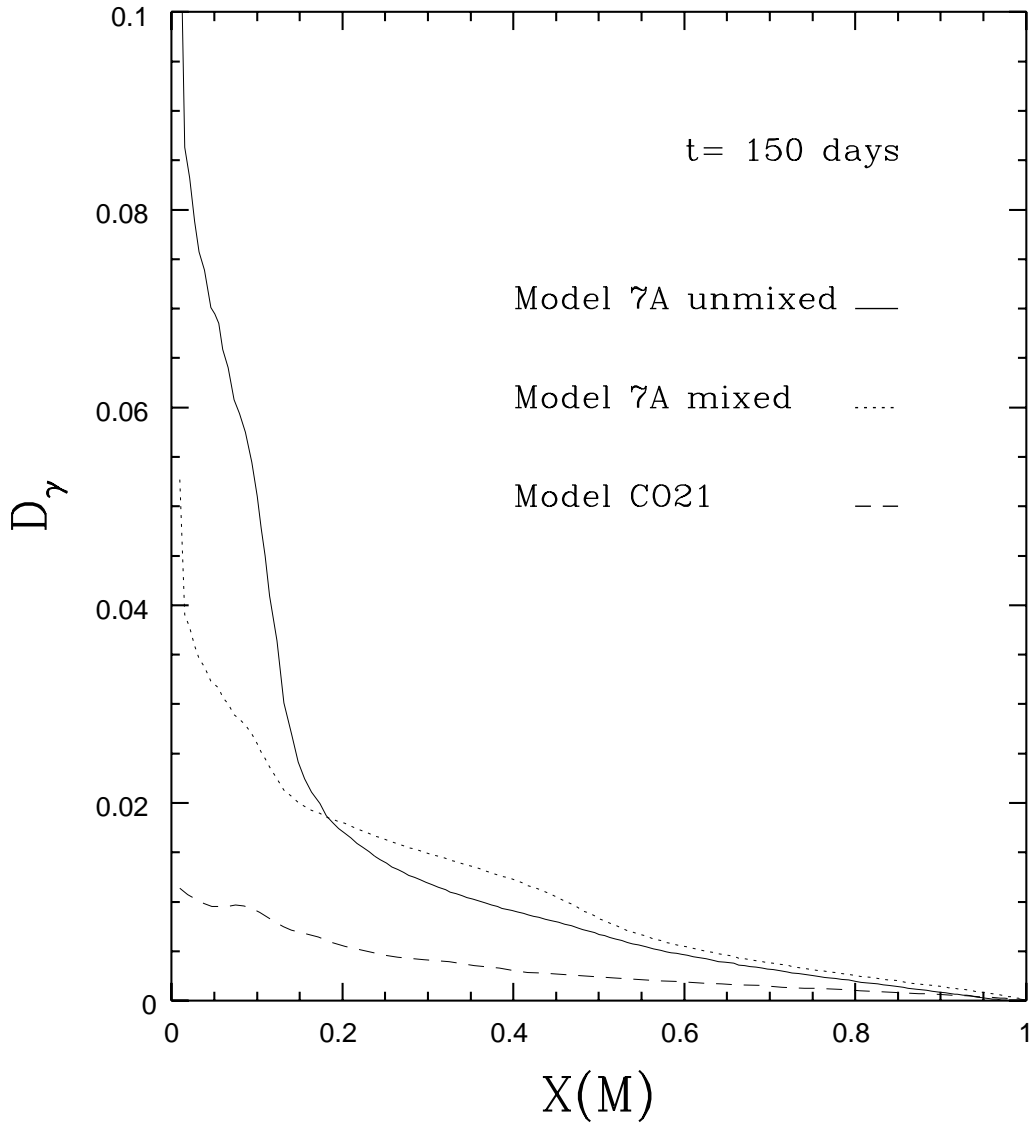}}
\nopagebreak[4]
\vspace{4mm}
\caption{Deposition of $\gamma$--rays for various models. Model 7A
 corresponds to the  
 30--40 M$\sb\odot$ main sequence Wolf--Rayet progenitor   
 for SNe Ibc by Woosley, Langer, \& Weaver 1996. Model 7A mixed is 
 a mixed version of Model 7A, which reproduces much better the spectra
 (Eastman \& Woosley 1997; Ruiz--Lapuente 1997). Model CO21 corresponds to 
 the less massive candidate to SNIbc of initial mass in the 
 range 10--20 M$\sb\odot$ which ends up as a 
 C+O star before the explosion (Nomoto et al. 1996).}
\end{figure}

\bigskip

\begin{figure}[hbtp]
\centerline{\epsfysize17cm \epsfbox{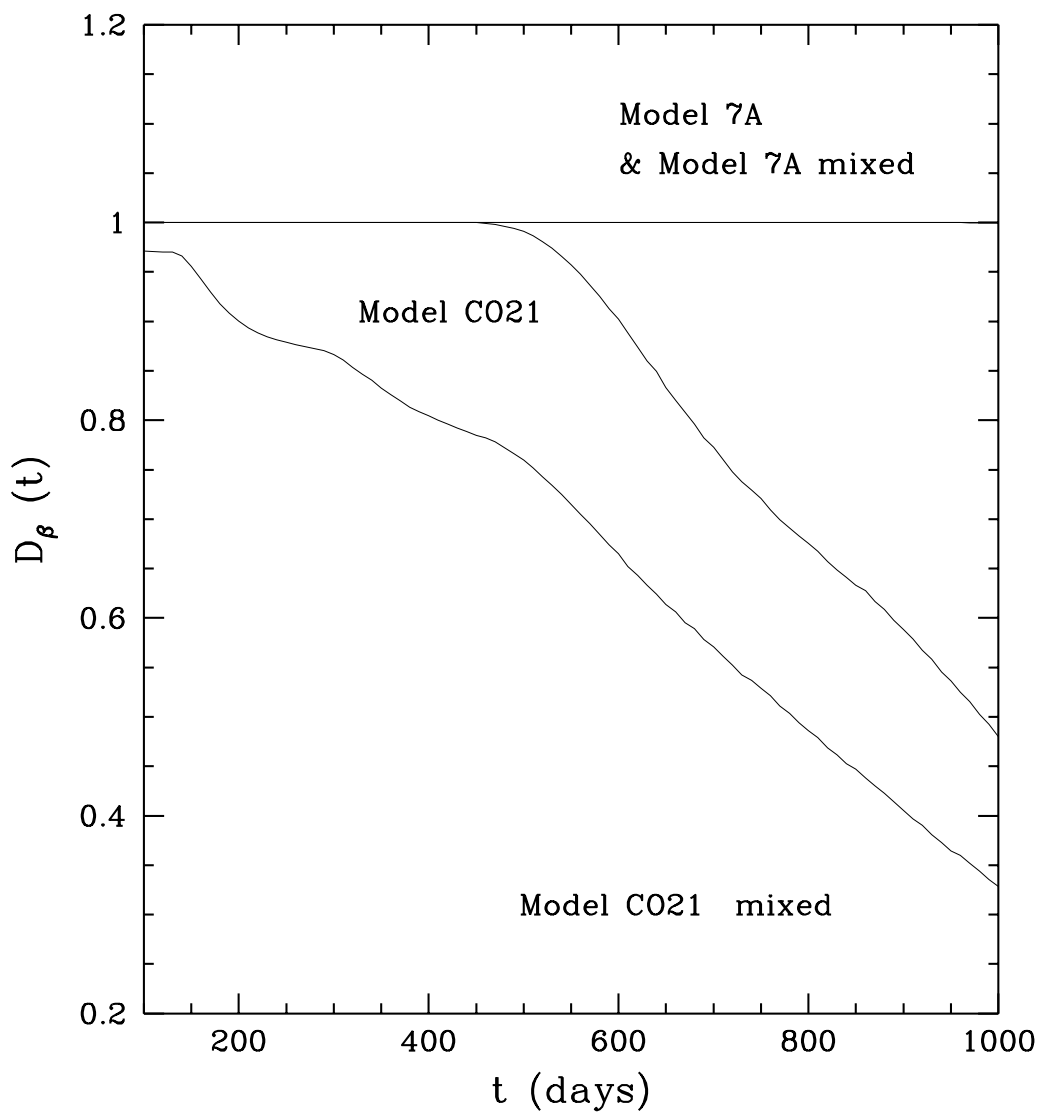}}
\nopagebreak[4]
\vspace{4mm}
\caption{Deposition of energy from e$^{+}$ for various models. Model 7A
 corresponds to the  
 30--40 M$\sb\odot$ main sequence Wolf--Rayet progenitor   
 for SNe Ibc by Woosley, Langer, \& Weaver 1996. Model 7A mixed is 
 a mixed version of Model 7A. Model CO21 corresponds to 
 the less massive candidate to SNIbc of initial mass in the 
 range 10--20 M$\sb\odot$ which ends up as a 
 C+O star before the explosion (Nomoto et al. 1996). Model CO21 mixed is
 a mixed version of Model CO21. Mixing is needed to reproduce the spectra.}
\end{figure}

\bigskip

\begin{figure}[hbtp]
\centerline{\epsfysize20cm\epsfbox{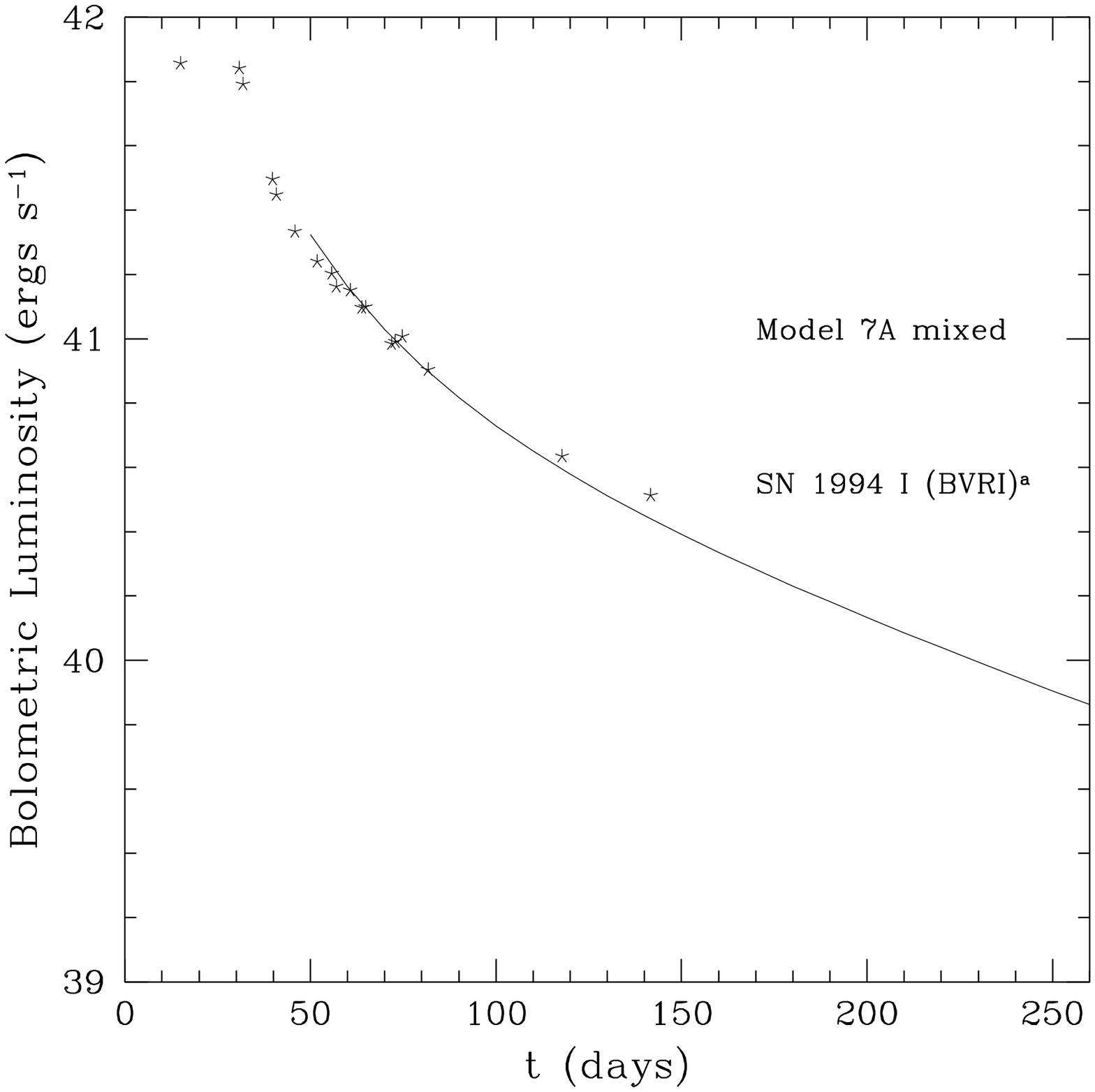}}
\nopagebreak[4]
\vspace{4mm}
\caption{Light curve of Model 7a mixed compared with
 the bolometric of SN 1994I by Richmond et al. (1996) taking
 as SN reddening E(B--V)=0.45.}
\end{figure}


\begin{thebibliography}{}
\bibitem{}
Arnett, W.D. 1969, Ap\&SS, 5, 180
\bibitem{}
Arnett, W.D. 1996, Supernovae and nucleosynthesis. Princeton series in
astrophysics, Princeton, NJ: Princeton University Press, p. 354
\bibitem{}
Axelrod, T. S. 1980, Ph D. Thesis, U. California at Santa Cruz 
\bibitem{}
Berger, M.J. \& Seltzer, S.M. 1964, Tables of Energy Losses \& Ranges of
Electrons \& Positrons (Washington, DC: NASA)
\bibitem{}
Blumenthal, G.B. \& Gould, R. J. 1970, Rev. Mod. Phys., 42, 237
\bibitem{}
Bravo, E. , Tornamb\'e, A., Dom\'\i nguez, I., \& Isern, J. 1996,   
A \& A, 306, 811 
\bibitem{}
Chan, K. \& Lingenfelter, R. E. 1993, ApJ, 403, 614
\bibitem{}
Colgate, S. A., Petschek, A.G., \& Kriese, J. T., 1980, ApJ, 237, L81 
\bibitem{}
Colgate, S. A. 1990. in Supernovae, ed. S. Woosley, (Springer--Verlag,
Berlin), 585
\bibitem{}
Colgate, S. A., Fryer, C. L., \& Hand, K. P. 1997. in Thermonuclear
Supernovae, ed. P. Ruiz--Lapuente, R. Canal \&  J. Isern. Kluwer Academic
Publishers, 273
\bibitem{}
Gould, R.J. 1972, Physica, 60, 145
\bibitem{}
Guessoum, N., Ramaty, R. \& Lingenfelter, R.E. 1991, ApJ 378, 170
\bibitem{}
Hamuy, M. et al. 1996a, AJ, 112, 2438
\bibitem{}
Hamuy, M. et al. 1996b, AJ, 112, 2408
\bibitem{}
Haymes, R.C. et al. 1975, ApJ, 201, 593
\bibitem{}
Heitler, w. 1954, The Quantum Theory of Radiation (3d. ed.; Oxford: Clarendon
Press)
\bibitem{}
Hernanz, M., Salaris, M., Isern, J. \& Jos\'e, J. 1997, in Thermonuclear
Supernovae, ed. P. Ruiz--Lapuente, R. Canal \& J. Isern. Kluwer Academic
Publishers, 167
\bibitem{}
Khokhlov, A. A\& A, L25 (1991)
\bibitem{}
Kirshner, R. P. 1997, private communication 
\bibitem{}
Kirshner, R.P \& Oke, J.B. 1975, ApJ, 200, 574
\bibitem{}
Kraichnan, R. H. 1976, J.Fluid. Mech, 77, 753 
\bibitem{}
Liebert, J. 1995 in Proceedings of the Cape Workshop on Magnetic Cataclysmic
Variables. ASP. Conference Series, Vol. 85. D. A. H. Buckley and B. Warner,
eds.
\bibitem{}
Lingenfelter, R.E. \& Ramaty, R. 1989, ApJ, 343, 686 
\bibitem{}
Livne, E. \& Arnett, D. 1995, ApJ 452, 62
\bibitem{}
Niemeyer, J. C. \& Hillebrandt, W. 1995, ApJ, 452, 769
\bibitem{}
Nomoto, K. 1982. ApJ 257, 780
\bibitem{}
Nomoto, K., Thielemann, F.--K. \& Yokoi, K. 1984, ApJ 286, 644 
\bibitem{}
Nomoto, K. 1995 (private communication)
\bibitem{}
Nomoto, K., Iwamoto, K., Young, T. R., Nakasato, N. \& Suzuki, T. 1997. in
Thermonuclear Supernovae, ed. P. Ruiz--Lapuente, R. Canal, \& J. Isern.
Kluwer Academic Publishers, 839
\bibitem{}
Porter, A. C., Dickinson, M., Stanford, S. A., Lada, E. A., Fuller, G. A.,
Myers, P.C. 1992, BAAS, 181, 7607
\bibitem{}
Ramaty, R. \& M\'esz\'aros, P. 1981. ApJ, 250, 384
\bibitem{}
Richmond, M. W., Van Dyk, S.D., Ho, W., Peng, C., Paik, Y., 
Treffers, R. R., \& Filippenko, A. V. 1996, AJ, 111, 327  
\bibitem{}
Riess, A.G., Press, W.H., \& Kirshner, R.P. 1996, ApJ 473, 88 
\bibitem{}
Roy, R.R. \& Reed, R.D. 1968, Interactions of Photons and Leptons with Matter
(New York: Academic)
\bibitem{}
Ruiz--Lapuente, P. et al. 1993, Nature, 365, 728
\bibitem{}
Ruiz--Lapuente, P. 1997 (in preparation)
\bibitem{}
Sandage, A., Saha, A., Tammann, G. A., Labhardt, L., Schwengeler, H.,
Panagia, N., \& Maccheto, F. D. 1994. ApJ 423, L13
\bibitem{}
Segr\'e, E. 1977, Nuclei and Particles (New York: Benjamin / Cummings)
\bibitem{}
Suntzeff, N. B. (1996). in IAU Colloq. 145, Supernovae and Supernova
Remnants, ed. R. McCray \& Z. W. Li (Cambridge, Cambridge University Press),
Cambridge, p. 41
\bibitem{}
Sutherland, P. G. \& Wheeler, J. C. 1984. ApJ 280, 282
\bibitem{}
Swartz, D.A., Sutherland, P. G. \& Harkness, R.P. 1995. ApJ 446, 766
\bibitem{}
Turatto, M., Benetti, S., Cappellaro, E., Danziger, I.J., Della Valle,
M., Gouiffes C., Mazzali, P. A., Patat, F. 1996, MNRAS, 283, 1 
\bibitem{}
Wallyn, P., et al. y 1993, ApJ 403, 621
\bibitem{}
Woosley, S. E. 1990, in Supernovae, ed. A. Petschek, (Springer--Verlag, 
Berlin), 182 
\bibitem{}
Woosley, S. E. \& Weaver, T. A. 1994, ApJ, 423, 371 
\bibitem{}
Woosley, S. E. 1997, in Thermonuclear Supernovae, ed. P. Ruiz--Lapuente, R.
Canal, \& J. Isern. Kluwer Academic Publishers, 313



\end{thebibliography}
\end{document}